\documentclass[a4paper,prl,reqno,superscriptaddress,twocolumn,showpacs,li]{revtex4-1}
\usepackage[utf8]{inputenc}
\usepackage{amsmath}
\usepackage{amsfonts}
\usepackage{amssymb}
\usepackage{makeidx}
\usepackage{graphicx}
\usepackage{color}
\usepackage{mathptmx}
\usepackage{bm}
\usepackage{changes,todonotes}

\def\bea{\begin{eqnarray}}
\def\eea{\end{eqnarray}}
\def\ba{\begin{array}}
\def\ea{\end{array}}
\def\n{\nonumber}

\def\la{\langle}
\def\ra{\rangle}

\def \dr{D_R}

\usepackage[utf8]{inputenc}
\usepackage{amsmath}
\usepackage{amsfonts}
\usepackage{amssymb}

\begin{document}

\title{Active Brownian Motion with Directional Reversals}
\author{Ion Santra}
\affiliation{Raman Research Institute, Bengaluru 560080, India}
\author{Urna Basu}
\affiliation{Raman Research Institute, Bengaluru 560080, India}
\affiliation{S. N. Bose National Centre for Basic Sciences, Kolkata 700106, India}
\author{Sanjib Sabhapandit}
\affiliation{Raman Research Institute, Bengaluru 560080, India}

\begin{abstract}
Active Brownian motion with intermittent direction reversals are common in a class of bacteria like {\it Myxococcus xanthus} and {\it Pseudomonas putida}. We show that, for such a motion in two dimensions, the presence of the two time scales set by the rotational diffusion constant $D_R$ and the reversal rate $\gamma$ gives rise to four distinct dynamical regimes: (I)  $t\ll \min (\gamma^{-1}, D_R^{-1}),$ (II) $\gamma^{-1}\ll t\ll D_R^{-1}$, (III) $D_R^{-1} \ll t \ll \gamma^{-1}$,  and (IV) $t\gg \max (\gamma^{-1}$, $D_R^{-1})$, showing distinct behaviors. We characterize these behaviors by analytically computing the position distribution and persistence exponents. The position distribution shows a crossover from a strongly non-diffusive and anisotropic behavior at short-times to a diffusive isotropic behavior via an intermediate regime (II) or (III). In regime (II), we show that, the position distribution along the direction orthogonal to the initial orientation is a function of the scaled variable $z\propto x_{\perp}/t$ with a non-trivial scaling function, $f(z)=(2\pi^3)^{-1/2}\Gamma(1/4+iz)\Gamma(1/4-iz)$. Furthermore, by computing the exact first-passage time distribution, we show that a novel persistence exponent $\alpha=1$ emerges due to the direction reversal in this regime. 
\end{abstract}

\maketitle
Active particles like bacteria, Janus colloids, and nanomotors are self-propelled, show persistent motion and manifest novel collective and single particle behavior~\cite{abp1,hydrodynamics1,separation2,active_computation,separation1,roadmapactive,Sriram,abp2,pressure}. Minimal statistical models capturing these features play a central role in the theoretical understanding of active matter~\cite{rtp2008,Bechinger,franosch1}. These models typically describe the overdamped motion of a particle with a constant speed $v_0$ along a stochastically evolving internal orientation.  The intrinsic nonequilibrium nature makes exact analytical treatment much more challenging, even for the minimal models, compared to their passive counterparts like the Brownian motion. Nevertheless,  analytical results for the position distribution and first-passage properties in certain situations have been obtained for two basic models---the so-called ``Run-and-tumble particle" (RTP)~\cite{1drtp,stadje,1drtptrap,maes1d,ion1,rtpddim} and the ``active Brownian particle" (ABP)~\cite{abp_potoski,sevilla2014,abp2018,abp2019,abp_potoski,abp2020,Majumdar2020,abp_polymer}. For RTP, the internal orientation changes by a finite amount via  an intermittent `tumbling' whereas, it  undergoes a rotational diffusion for ABP. These models successfully describe dynamics of bacteria like {\it E. coli} and `catalytic-swimmers'~\cite{ecoli,abp0,abpexp}.

Many microorganisms---such as \emph{Myxococcus xanthus}~\cite{xanthus1,xanthus2,xanthus3,xanthus4}, \emph{Pseudomonas putida}~\cite{putida1,putida2},  {\it Pseudoalteromonas haloplanktis} and {\it Shewanella putrefaciens}~\cite{marine1, marine2}, and   {\it Pseudomonas citronellolis}~\cite{monoperitrichous}---however, show a distinctly different dynamics. They undergo intermittent directional reversals, in addition to an ABP like motion. The origin of such reversals is different in different organisms, eg., internal protein oscillations reverse the cell polarity which causes the directional reversal in \emph{Myxococcus xanthus}~\cite{xanthus1,xanthus3} while a reversal of swimming direction occurs due to the reversal in the rotation direction of polar flagella in \emph{Pseudomonas putida} ~\cite{putida1,putida2}. The addition of the drastic reversal dynamics to the rotational diffusion gives rise to a host of emergent collective phenomena including fruiting body formation\cite{xanthus2}, generation of rippling patterns~\cite{xanthusagg1} and  accordion waves~\cite{xanthusagg2}.

Despite the widespread appearance of this direction reversing active Brownian particles (DRABP), a theoretical understanding of it is still lacking---even at the level of single particle position distribution. Another relevant observable for active particles like bacteria is the  first-passage time~\cite{redner,fprev2} to reach a particular target such as food source, weak spot of the host or toxins.  For example, certain starvation induced complex processes have been seen in {\it Myxococcus xanthus}~\cite{xanthus2} and {\it Pseudomonas putida}~\cite{starvation_putida}, which in turn would depend on the first-passage properties. Again, no theoretical results are available for the  first-passage statistics of the DRABP. In this Letter we obtain exact analytical results for the position distribution and persistence exponents describing the power-law decay of the survival probability, thus providing a  comprehensive theoretical understanding.

\begin{figure}[t]
\includegraphics[width=0.9\hsize]{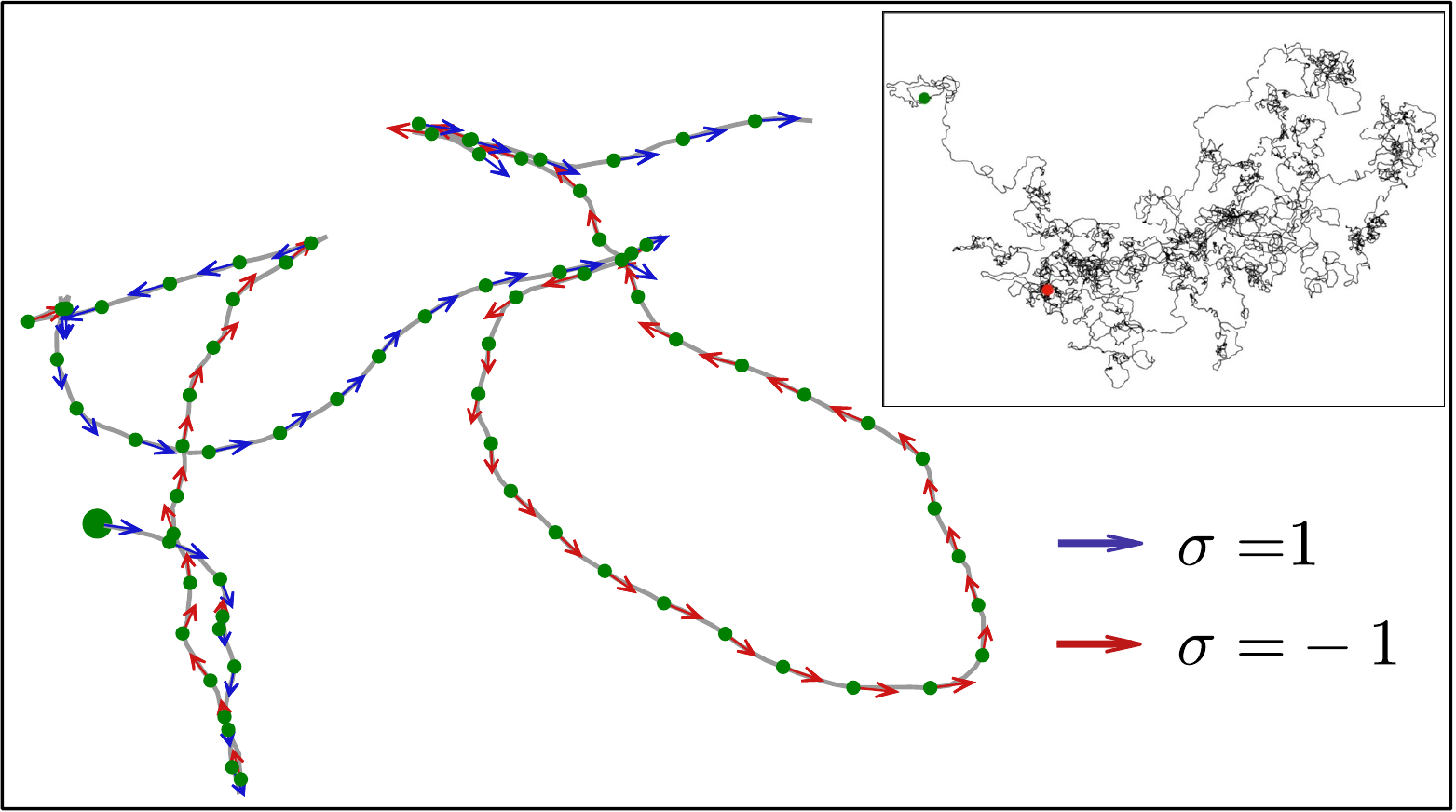}
\caption{A typical trajectory of a DRABP generated by discretizing  \eqref{eq:model}, where in a small interval $\Delta t,$ the particle reverses the direction with probability $\gamma \Delta t$ and with probability $1-\gamma \Delta t$ performs an ABP: $\{\Delta x(t),\Delta y(t)\}=v_0 \sigma(t)\Delta t\{\cos\theta(t),\sin\theta(t)\};\,\Delta \theta(t)=\sqrt{2 D_R \Delta t}\, \chi$, where $\chi$ is drawn from a standard normal distribution. The arrows indicate the instantaneous velocity vectors. The inset shows a long-time trajectory [which resembles a Brownian trajectory] with the two end-points marked. See~\cite{SM} for an animation.}\label{fig:traj}
\end{figure}

\begin{figure*}[t]
\hspace*{-0.5cm}\includegraphics[width=17.6 cm]{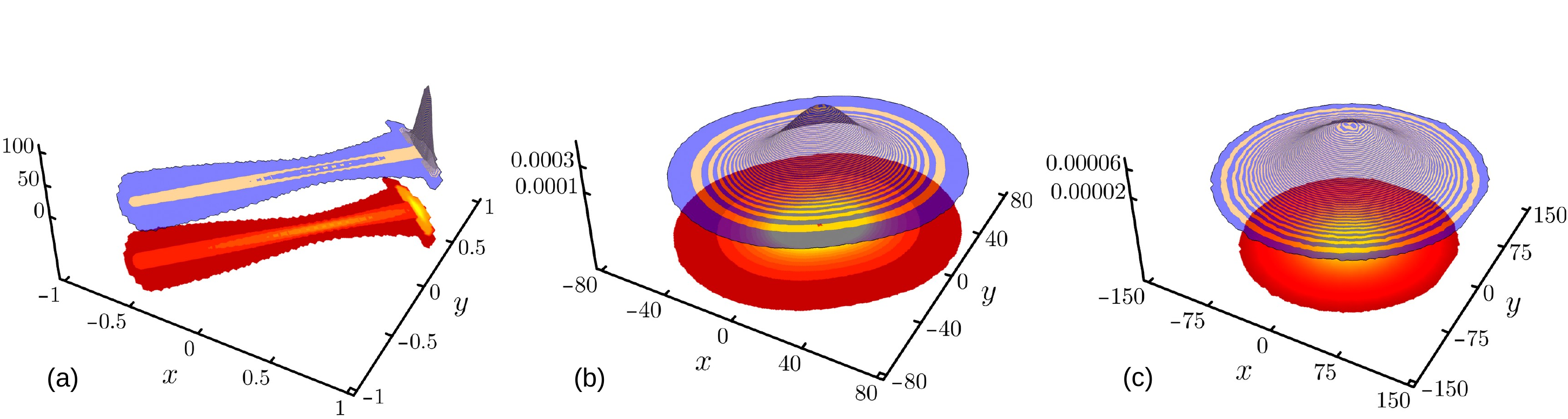}
\caption{Dynamical evolution of the position distribution $P(x,y,t)$ for the case $\gamma > \dr$   obtained from numerical simulations. Here we have taken  $ \gamma=0.1$, $D_R=0.01$ and initial orientation $\theta_0=\pi/4.$  The panels (a), (b) and (c) correspond to $t=1$ [regime (I)], $t=50$ [regime (II)] and $t=200$ [regime (IV)], respectively.  The strong anisotropy in regime (I) persists in the intermediate regime (II) eventually disappearing at large times (IV). See~\cite{SM} for a movie. For $\gamma < \dr$ the intermediate regime (b) is replaced by regime (III) which looks similar to (c).}
\label{fig:map}
\end{figure*}  

In two dimensions, the position $\bm{x}= (x,y)$ and orientation $\theta$ of a DRABP evolve according to the Langevin equations,
\begin{subequations}
\label{eq:model}
\begin{align}
\label{eq:x}
\dot{x}(t)&= v_0\, \sigma(t) \cos \theta(t)\equiv \zeta_x(t),\\
\label{eq:y}
\dot{y}(t)&= v_0 \,\sigma(t)\sin \theta(t)\equiv\zeta_y(t), \\
 \label{eq:theta}
\dot{\theta}(t)&=\sqrt{2D_R}~\eta (t),
\end{align}
\end{subequations}
where $\dr$ is the rotational diffusion coefficient, $\eta(t)$ is a Gaussian white noise with $\langle \eta(t)\rangle =0$ and $\langle \eta(t) \eta(t')\rangle =\delta(t-t').$ The dichotomous noise $\sigma(t)$ alternates between $\pm 1$ at a constant rate $\gamma,$ triggering the direction reversal [see Fig.~\ref{fig:traj} for a typical trajectory]. 

\begin{figure}
\includegraphics[width=\hsize]{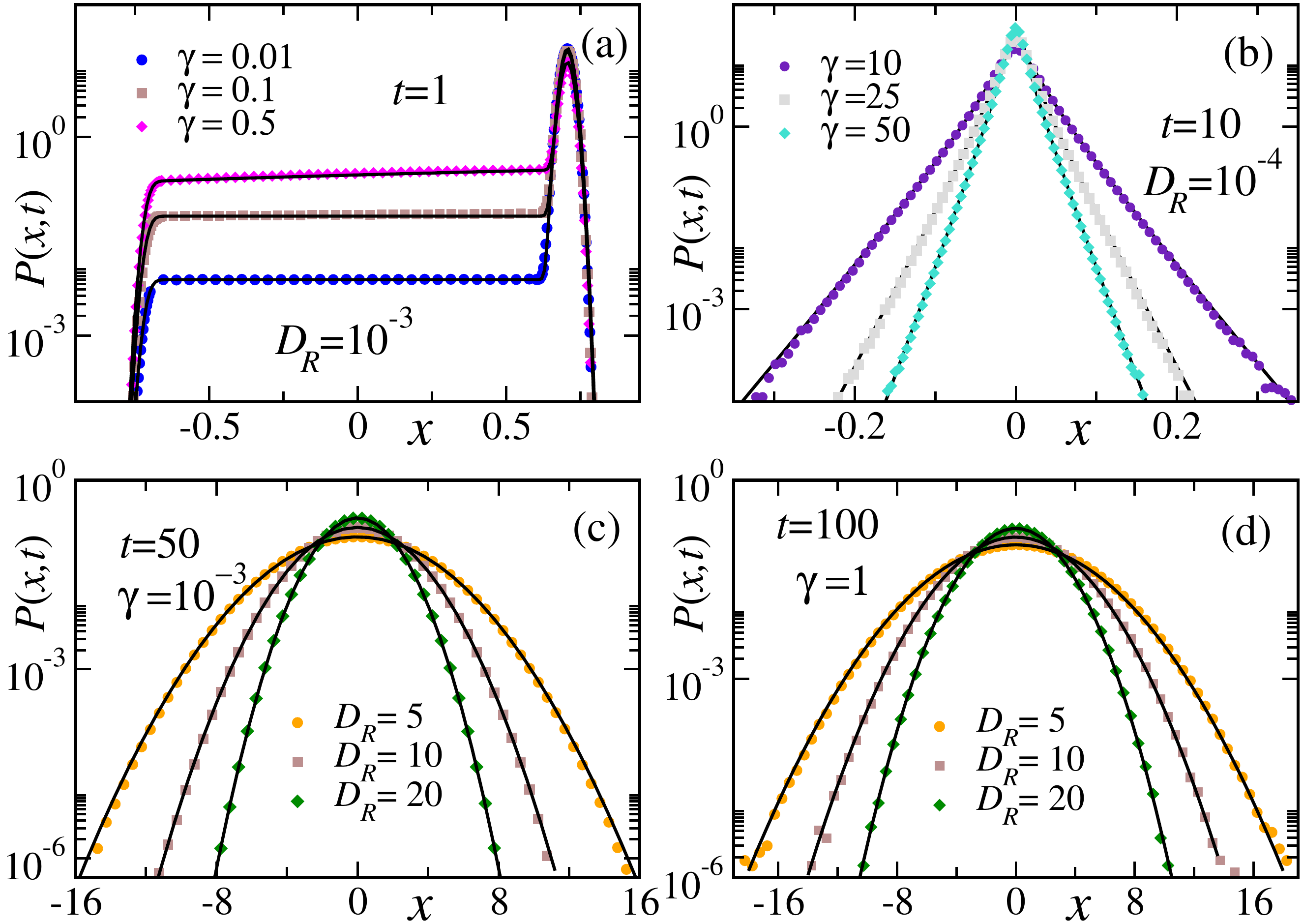}
\caption{Marginal position distribution $P(x,t)$ in the different dynamical regimes: (a) $t \ll \min(D_R^{-1}, \gamma^{-1})$, (b) $\gamma^{-1} \ll t \ll D_R^{-1}$, (c)  $D_R^{-1} \ll t \ll \gamma^{-1}$, (d) $t \gg \max(D_R^{-1}, \gamma^{-1})$. 
The symbols are from numerical simulations while solid black lines correspond to the analytical predictions given by Eqs.~\eqref{eq:Pxt_short} [up to $n=2$], ~\eqref{intermediate_scaling},~\eqref{intermed_D>g} and ~\eqref{larget} for (a)-(d) respectively. Here $v_0=1$ and we have used initial orientation $\theta_0= \pi/4$ for (a), (c) and (d) and $\theta_0=\pi/2$ for (b).} \label{fig:pxt}
\end{figure}

 In this Letter, we show that the presence of the two time-scales, $\dr^{-1}$ and  $\gamma^{-1}$, gives rise to four distinct dynamical regimes: (I)  $t\ll \min (\gamma^{-1}, D_R^{-1}),$ (II) $\gamma^{-1}\ll t\ll D_R^{-1}$, (III) $D_R^{-1} \ll t \ll \gamma^{-1}$,  and (IV) $t\gg \max (\gamma^{-1}$, $D_R^{-1})$, each characterized by a different dynamical behavior. Indeed, for {\it Myxococcus xanthus}, the well-separated time-scales ($\gamma^{-1}\simeq 10^2$~s and $~D_R^{-1}\simeq 10^6$~s ~\cite{xanthustime}) make the regimes (I), (II) and (IV) experimentally accessible.
%
%
  We find that  the position distribution shows a crossover from a strongly non-diffusive and anisotropic behavior at short-times to an eventual isotropic diffusive behavior via an intermediate regime (II) or (III) whose behaviors are very different. In regime  (I), starting from the origin with a fixed orientation $\theta_0$, the position distribution of a DRABP is strongly anisotropic and shows a plateau-like structure around the origin accompanied by a single peak near $v_0 t$ along $\theta_0$ [Figs.~\ref{fig:map}(a) and \ref{fig:pxt}(a)]. For $\gamma>D_R$  the anisotropy persists in the intermediate time regime (II), however, with a peak at the origin [Fig.~\ref{fig:map}(b)]. In particular, we show that, the position distribution along the direction orthogonal to the initial orientation has a scaling form 
\bea
P(x_\perp,t) =   \frac {1}{v_0t}\sqrt{\frac \gamma{8 \dr}} f\left(\frac {x_\perp}{v_0t}\sqrt{\frac \gamma{8 \dr}}\right),
\label{intermediate_scaling}
\eea
with an exact non-trivial scaling function,
\bea
f(z) =\frac 1 {\sqrt {2 \pi^3}} \Gamma\left(\frac 14 + i z \right) \Gamma\left(\frac 14 - i z\right), \label{intermediate_exact}
\eea
where $\Gamma(z)$ is the gamma function. The tails of  $f(z)$ decay exponentially [Fig.~\ref{fig:pxt}(b)]. Regime (III) appears for  $D_R>\gamma$ where  the distribution is Gaussian [Fig.~\ref{fig:pxt}(c)] with the variance $v_0^2/(2 \dr)$. The distribution is also a Gaussian in the late-time regime (IV), albeit with a different variance $v_0^2/[2(2\gamma+\dr)]$ [Figs.~\ref{fig:map}(c) and \ref{fig:pxt}(d)].

The persistence property also shows distinct  behaviors in the four dynamical regimes (I)--(IV). We show that the directions parallel and orthogonal to the initial orientation are characterized by different persistence exponents, $\alpha_{\parallel}$ and $\alpha_{\perp}$ respectively, which are summarized in Table~\ref{survxtab}. The most noteworthy is a new persistence exponent $\alpha_{\perp}=1$ in the intermediate regime (II), emerging due to the presence of the direction reversal. In particular, in the limit $\gamma\to\infty$ and $D_R\to 0$, the first-passage time distribution for the perpendicular component has the scaling form 
\bea
F_{\perp}(t;x_{\perp0})=\frac{x_{\perp0}\, \sqrt{2} \gamma^{3/2}}{v_0^3\, t^2\sqrt {\dr}} f\left(\frac {x_{\perp0}}{v_0t}\sqrt{\frac \gamma{8 \dr}}\right),
\label{fpt_exact}
\eea
where $f(z)$ is given by Eq.~\eqref{intermediate_exact} and $x_{\perp0}$ is the initial position. In fact, for $\gamma > \dr$,  we find that, $\alpha_\perp$ shows a non-monotonic behavior--- $\alpha_\perp=1/4$ at short-times (I), which crosses over to $\alpha_\perp=1$ in the regime (II), and finally reaches the Brownian value $\alpha_\perp=1/2$ at late times (IV). 

{\it Position distribution.---} We begin by considering the position distribution. The correlated nature of the effective noises $\zeta_{x,y}(t)$ in Eq.~\eqref{eq:model} makes the dynamics non-diffusive and anisotropic at short-times [see Sec. I of Supplemental Material~\cite{SM} for details]. In the following we first consider the two extreme regimes (I) and (IV) before coming to the intermediate regimes (II) and (III). We set $x(0)=y(0)=0$ without any loss of generality.

 {\it Short-time regime} (I): Starting from the initial orientation $\theta(0) = \theta_0$ and $\sigma(0)=1,$ for $t \ll D_R^{-1}$ the effective noises can be approximated as,
\begin{subequations}
\label{smallD_langevin}
\bea
\zeta_x(t)\approx v_0\sigma(t)(\cos \theta_0 - \phi(t) \sin \theta_0 ), \label{eq:x_short_t} \\
\zeta_y(t)\approx v_0 \sigma(t)(\sin \theta_0 + \phi(t)\cos \theta_0 ), \label{eq:y_short_t}
\eea \end{subequations}
where $\phi(t) = \sqrt{2 D_R} \int_0^t \eta(s)\,ds$ denotes a standard Brownian motion. Here we have approximated $\cos \phi(t) \simeq 1$ and $\sin \phi(t) \simeq \phi(t)$ for  $t\ll D_R^{-1}$ as $\phi(t) \sim \sqrt{\dr t} \ll 1$~\cite{abp2018}. 

To obtain the marginal position distributions $P(x,t)$ and $P(y,t)$ corresponding to Eqs.~\eqref{smallD_langevin}, we adopt a trajectory based approach. The trajectory of the DRABP over a time-interval $[0,t]$ can be divided into $(n+1)$ intervals, punctuated by $n$ direction reversals; $\sigma$ remains constant between two consecutive reversals. We show that, for  a specific sequence  of intervals with duration $\{ s_1,s_2,\dotsc,s_{n+1} \},$ the distribution of the final position $x$ is a Gaussian with mean $\cos \theta_0 \sum_{i=1}^{n+1} (-1)^i s_i$ and variance $b_n\, \sin^2 \theta_0 $, where
\bea
b_n = 2 \dr \sum_{i=1}^{n+1} \Bigg[ \sum_{j=1}^{i-1} (-1)^{i+j} s_i s_j(t_j+t_{j-1}) + \frac{s_i^2}3 (t_i+2t_{i-1}) \Bigg].\n
\eea
Here $t_i = \sum_{j=1}^{i} s_j$  with $t_0=0$ and $t_{n+1}=t.$ The position distribution is then obtained by  taking weighted contributions from all such trajectories. Skipping details (Sec.~III of~\cite{SM}), we get,
\begin{align}
P(x,t)=&\frac {e^{-\gamma t}}{v_0\sin \theta_0 \sqrt{2 \pi}} \sum_{n=0}^{\infty} \gamma^n \int_0^t \prod_{i=1}^{n+1} ds_i \frac{\delta(t - \sum_{i=1}^{n+1}s_i)}{\sqrt{b_n}} \n\\
&\times \exp{\left[ - \frac{(x + v_0\cos\theta_0 \sum_{i=1}^{n+1}(-1)^i s_i)^2}{2v_0^2 \sin^2 \theta_0 b_n} \right]},\label{eq:Pxt_short} 
\end{align}
where each term corresponds to trajectories with a fixed number $n$ of reversals. We also obtain the $y$-marginal distribution $P(y,t)$ in the same manner, whose explicit form is given in~\cite{SM}.

Equation~\eqref{eq:Pxt_short} provides the exact time-dependent marginal distribution  for the process \eqref{smallD_langevin}.  Even though the infinite series cannot be summed explicitly to obtain a closed-form expression, it can be systematically evaluated numerically to obtain $P(x,t)$   for arbitrary  $\gamma$ and $t.$ In fact, for $t \lesssim \gamma^{-1},$ it suffices to consider the first few terms to get a reasonably good estimate of the marginal distributions~\cite{SM}. Figure~\ref{fig:pxt}(a)  compares this estimate, evaluated up to $n=2$ terms, with $P(x,t)$ obtained from numerical simulations. Clearly, this perturbative approach is extremely successful in accurately predicting the characteristic shape of the distribution, with a wide plateau  near the origin and a peak near $x= v_0 t$, in this short-time regime~(I).

Physically, the peak in the distribution is a manifestation of the  ABP nature of the motion---the $n=0$ term, corresponding to the no reversal case, correctly predicts the peak. The emergence of the plateau, however, is a direct consequence of the reversal events---for $t \ll D_R^{-1},$ the orientation $\theta$ evolves slowly, and the dynamics can be thought of as a one-dimensional RTP with an effective velocity $v_0 \cos \theta_0.$ Now, for small values of $\gamma,$ the trajectories with a single flip contribute a constant value (the plateau) $ {\gamma e^{-\gamma t}} /(2 v_0 \cos \theta_0)$. This agrees well with the exact result [Eq.~\eqref{eq:Pxt_short}] to leading order in $\gamma.$ Interestingly, such a plateau with a boundary-peak has been observed for motile bacteria in emulsion droplets~\cite{plateau_experiment}.

The anisotropic nature of the distribution in this short-time regime is a direct artifact of the fixed initial orientation. If, instead, the initial orientation is chosen uniformly, the position distribution becomes isotropic and an additional peak emerges at the origin (Fig.~3 of~\cite{SM}).

{\it Long-time regime} (IV): For a given time $t$, mathematically, this regime can be accessed by taking both $D_R$ and $\gamma$ large ($\gg t^{-1}$). For large $D_R$ and arbitrary $\gamma$, the effective-noise autocorrelation becomes~(Sec.~II of~\cite{SM}), $\la \zeta_{a}(t)\zeta_{b}(t')\ra\approx 2D_{\text{eff}}\delta_{a,b}\,[D_R+2\gamma]\exp\bigl(-[D_R+2\gamma]|t-t'|\bigr)$ where $2D_{\text{eff}}=v_0^2/{(\dr+2\gamma)}$. Thus in the limit $D_R,\gamma\to\infty$, it tends to $2D_{\text{eff}}\delta_{a,b}\delta(t-t')$ which results in the isotropic Gaussian distribution, 
\bea
P(x,y,t)&\approx& \frac{1}{ 2D_{\text{eff}}t}G\left(\frac{x}{ \sqrt{2D_{\text{eff}}t}},\frac{y}{ \sqrt{2D_{\text{eff}}t}}\right),
\label{larget}
\eea
with $G(\tilde x, \tilde y)= e^{-(\tilde x^2+\tilde y^2)/2}/(2\pi)$. The corresponding $x$-marginal distribution (which, obviously, is also a Gaussian) is plotted in Fig.~\ref{fig:pxt}(d) along with the data from numerical simulations; an excellent agreement validates our prediction.

{\it Intermediate-time regime} (III): This regime corresponds to $D_R\gg t^{-1}\gg \gamma$, where $\la \zeta_a(t)\zeta_b(t') \ra\to (v_0^2/D_R)\delta_{a,b}\,\delta(t-t').$ Therefore, the typical position distribution is again Gaussian with the width $v_0\sqrt{t/D_R}$, 
\bea
P(x,y,t) \approx \frac{\dr}{v_0^2 t} \, G\left(\frac{x }{v_0 \sqrt{t/\dr}},\frac{y }{v_0 \sqrt{t/\dr}}\right),~ \label{intermed_D>g} 
\eea
with $G(\tilde x, \tilde y)= e^{-(\tilde x^2+\tilde y^2)/2}/(2\pi).$ Note that this result is same as in the case of ABP for $t\gg D_R^{-1}$~\cite{abp2018} ---adding directional reversal does not change the physical scenario in this regime. 
 We validate this prediction with numerical simulations in Fig.~\ref{fig:pxt}(c).

{\it Intermediate-time regime} (II):  The correlated noise leads to an intriguing behavior in this regime $\gamma^{-1} \ll t \ll D_R^{-1}.$ For $t\gg\gamma^{-1},$  
the frequent reversals lead to a Gaussian white noise $\xi(t)$  with zero-mean and the correlator $\la\xi(t)\xi(t')\ra= \gamma^{-1}\delta(t-t')$. Thus, for $t\ll D_R^{-1}$, from Eqs.~\eqref{smallD_langevin}, the effective noises can be approximated as,
\begin{subequations}
\label{intermed_langevin}
\bea
\zeta_x(t)&\approx & v_0 \,\xi(t)(\cos \theta_0 - \phi(t) \sin \theta_0 ),\label{eq:x_intermed_langevin}\\
\zeta_y(t)&\approx & v_0 \,\xi(t)(\sin \theta_0 + \phi(t) \cos \theta_0).\label{eq:y_intermed_langevin}
\eea 
\end{subequations}
 Equations~\eqref{intermed_langevin} describe a Brownian motion with stochastically evolving diffusion coefficients. Some specific versions of such models have been studied recently~\cite{diffusing_diff} in a different context.

The Gaussian nature of $\xi(t)$, for a fixed $\{\phi(s)\}$ trajectory, allows us to evaluate the characteristic function, $\la e^{i \bm{k} \cdot \bm{x}} \ra = \left\la \exp{\left[- \frac {1}{2} \bm{k}^\text{T} \, \bm{\Sigma}(t) \, \bm{k} \right]} \right\ra_\phi,$ where $\bm{k} = \left({k_x, k_y} \right)^{\text{T}}$ and ${\bm \Sigma}(t)$ is the correlation matrix whose explicit form is given in Sec. IV of \cite{SM}. The subscript $\phi$ denotes averaging over the Brownian paths $\{\phi(s)\}$, which can be performed using path integral approach. This yields~\cite{SM},
\bea
\la e^{i \bm{k} \cdot \bm{x}} \ra = \frac{1}{\sqrt{\cosh\omega t}}\exp\left[-\frac{\omega \tanh \omega t}{4D_R} \left(\frac{k_x  + k_y\tan \theta_0}{k_y - k_x \tan \theta_0}\right)^2  \right],\nonumber\\
\label{intermediate_exactGF}
\eea
where $\omega = v_0\sqrt{{2D_R}/{\gamma}}\, (k_y\cos\theta_0-k_x\sin\theta_0).$ 
Of particular interest are the distributions along and orthogonal to the initial orientation, denoted by $x_{\parallel}$ and $x_{\perp}$ respectively. Setting $\theta_0=0$ gives $x_{\perp}\equiv y$ and $x_{\parallel}\equiv x$. 

Putting $k_x=0,\,k_y=k$ in Eq.~\eqref{intermediate_exactGF}, yields $\la e^{ikx_{\perp}}\ra=\left[\cosh \left(v_0 kt \sqrt{{2D_R}/{\gamma}}\right) \right]^{-\frac 12}$, which leads to the non-trivial distribution announced in Eqs.~\eqref{intermediate_scaling}-\eqref{intermediate_exact} for $x_{\perp}$.
Figure~\ref{fig:pxt}(b) shows an excellent agreement between  Eq.~\eqref{intermediate_exact} and the numerical simulations. The tails of the distribution decay as $\sim \exp[-\pi|x_{\perp}|/(2\sqrt{2}\ell)]$, where the charachteristic length-scale $\ell=v_0 t\sqrt{\dr/\gamma}$ is the root-mean-square displacement (Eq. 14 in SM) in regime (II). It is evident from the ballistic scaling form Eq.~\eqref{intermediate_scaling} that the variance $ \propto t^2$ (Eq.~(14) in \cite{SM}).
On the other hand, $k_y=0,\, k_x=k$ gives $\la e^{ikx_{\parallel}}\ra = e^{-v_0^2k^2\,t /(2\gamma)}$, which leads to a Gaussian distribution with a variance $v_0^2 t/\gamma$, indicating diffusive fluctuations. This drastically different nature of the fluctuations for $x_{\perp}$ and $x_{\parallel}$ leads to the anisotropic distribution seen in Fig.~\ref{fig:map} (b).

{\it  First-passage properties.---} We next consider the survival probability which is the cumulative distribution of the first-passage time. We set $\theta_0=0$ so  that $x_{\parallel}=x$ and $x_{\perp}=y$. Let $S_y(t;y_0)$ denote the probability  that, starting from some arbitrary position $y(0)=y_0$ the $y$-component of the position has not crossed the $y=0$ line up to time $t$; $S_x(t;x_0)$ is defined similarly.  
 
\begin{figure}[t]
\includegraphics[width=8.8 cm]{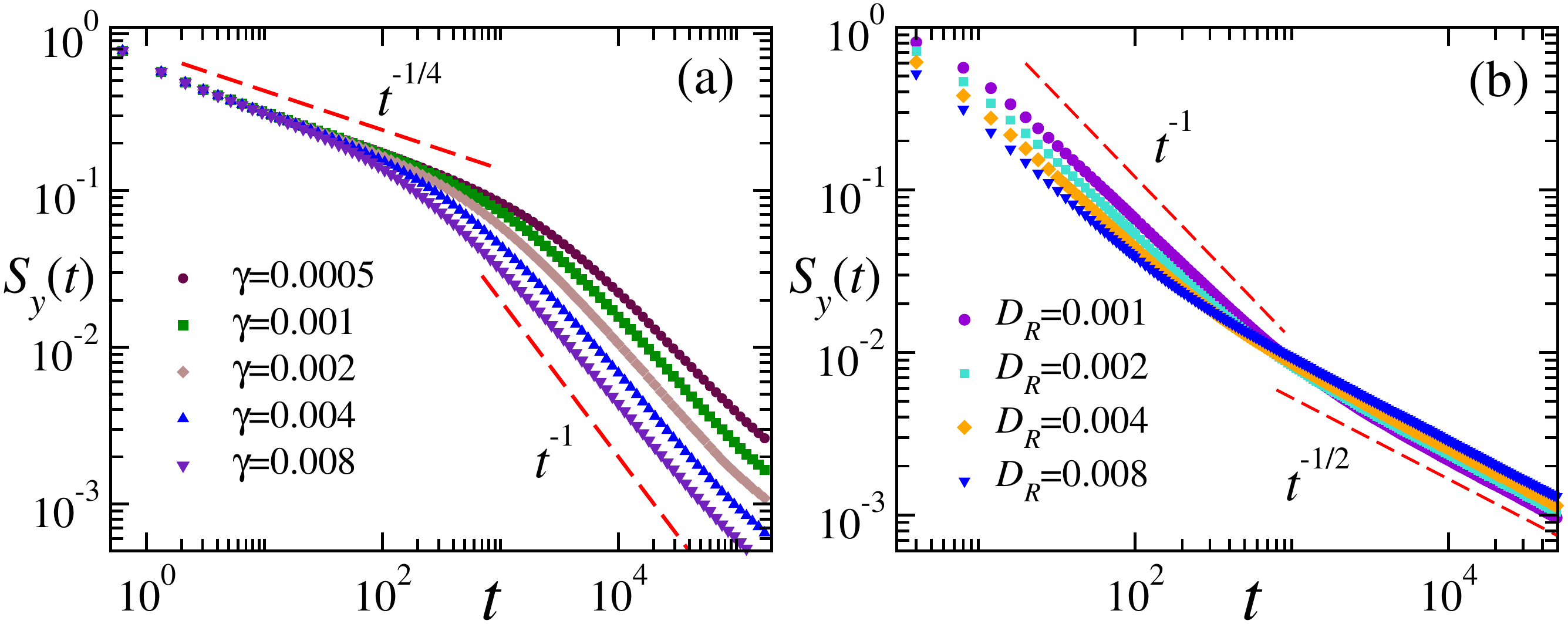} 
\caption{$S_y(t)$ {\it vs} $t$ for $\gamma>D_R$: (a) shows the crossover from $\alpha_y=1/4$ (I) to $\alpha_y=1$ (II) for a fixed $D_R=10^{-5}$ and $y_0=0.001.$ (b) shows the crossover from $\alpha_y=1$ (II) to $\alpha_y=1/2$ (IV) for $\gamma=1$ and $y_0=0.1$.}
\label{fig:surv}
\end{figure}

The most interesting scenario appears for $\gamma>D_R$ where  $S_{y}(t)$ shows three distinct persistence behaviors  in the three different dynamical regimes. For  $\theta_0=0$, Eq.~\eqref{eq:y_short_t} leads to $\dot{y}=v_0\sigma(t)\phi(t)$. Now, for $t\ll\gamma^{-1}$, this can be approximated as a random acceleration process $\dot{y}\simeq v_0\sigma(0)\phi(t)$, for which the persistence exponent is $1/4$~\cite{rap1,rap2}. Therefore, in the short-time regime (I), we expect $\alpha_y=1/4$, which is verified in Fig.~\ref{fig:surv}(a) using numerical simulations. 
%
%
 
In the intermediate time regime (II) the effective $y$-dynamics   [see \eqref{eq:y_intermed_langevin}] becomes $\dot{y}=v_0 \xi(t)\phi(t)$ for $\theta_0=0.$ 
We compute the survival probability by solving the corresponding Fokker Planck equation with an absorbing boundary condition at $y=0$. 
This leads to  a new persistence exponent $\alpha_y=1$ in the context of active particles. In fact, in the limit $\gamma\to\infty$ and $\dr\to 0 $, we find the exact first-passage time distribution  Eq.~\eqref{fpt_exact}, which we verify using numerical simulations in Sec.~V of \cite{SM}. This result is consistent with the recently obtained first-passage behavior for the diffusing diffusivity model~\cite{fpt_DD}. We show the crossover from $\alpha_y=1/4$ to $\alpha_y=1$ near $t\sim \gamma^{-1}$ using numerical simulations in  Fig.~\ref{fig:surv}(a).

In the large time regime (IV), as discussed earlier, the particle behaves like an ordinary diffusion process with an effective diffusion constant. Consequently, the survival probability decays with the Brownian exponent $\alpha_y=1/2$ as seen from the numerical simulations in Fig.~\ref{fig:surv}(b). The crossover from $\alpha_y=1$ to $\alpha_y=1/2$ occurs around $t\sim D_R^{-1}$.

If $D_R>\gamma$, we effectively see two distinct exponents $\alpha_y=1/4$ in the short-time regime (I), which crosses over to $\alpha_y=1/2$ in regime (III) and remains the same for large times (IV). We also study $S_x(t)$  which shows Brownian behavior at all times except in regime (I), where $\alpha_x=0$ due to the fact that the particle always survives for $t\ll\gamma^{-1}$; see Fig.~6 in \cite{SM}. A summary of the
exponents in all the regimes is provided in Table~\ref{survxtab}. \\

\begin{table}[t]
\begin{tabular}{|c|| p{1 cm}| p{1 cm}| p{1 cm}| p{1 cm}|}
\hline
\text{\shortstack{~ Persistence ~\\ Exponent }} & \;\quad I & \quad II & ~~III & \quad IV \\
\hline
$\alpha_{\perp}$ & ~~$1/4$ & ~~~$1$ & ~~$1/2$ & ~~$1/2$\\
\hline
$\alpha_{\parallel}$ & ~~~$0 $ & ~~$1/2$ & ~~$1/2$ & ~~$1/2$ \\
\hline
\end{tabular}
\caption{Persistence exponents for the $x_{\perp}$ and $x_{\parallel}$ components of the DRABP as defined by decay of the survival probability $S(t) \sim t^{-\alpha}$ in the different dynamical regimes: (I) $t \ll \min(D_R^{-1},\gamma^{-1}),$  (II) $\gamma^{-1} \ll t \ll D_R^{-1},$ (III) $D_R^{-1} \ll t \ll \gamma^{-1},$ and (IV) $t\gg \max(\gamma^{-1},D_R^{-1}).$ }\label{survxtab}
\end{table}

In conclusion, we provide a comprehensive analytical understanding of the DRABP that models a wide range of bacterial motion. The DRABP shows many novel features  --- the presence  of the direction reversal along with rotational diffusion gives rise to four distinct dynamical regimes each of which corresponds to a different position distribution and persistence exponent. In particular, we find that, the position distribution in the short-time and intermediate-time regimes have certain unique features, very different from ordinary ABP and RTP. The short-time regime is characterized by the emergence of a plateau.  The intermediate regime $\gamma^{-1} \ll t \ll \dr^{-1}$ shows a unique scaling behavior [see Eqs.~\eqref{intermediate_scaling}-\eqref{intermediate_exact}]. We also  find a novel persistence exponent $\alpha=1$, which have not been seen in active motions so far, in the same regime.

Our work opens up the possibility to explore a wide range of problems in the study of active motions, both experimentally and theoretically. Our results for  the position distribution may be verified in experiments with dilute solutions of bacteria~\cite{xanthus2,putida2} or artificial active colloids~\cite{Verman2020}. Moreover, it would be really interesting to observe the non-monotonic persistence exponent behavior from single particle tracking.  
In this Letter we have considered the direction reversal to be a Poisson process. To study the effect of other reversal-time distributions, like Gamma-distributions~\cite{grossman1,detcheverry} or a power-law, is a challenging open question. It would also be interesting to study the effect of interaction on DRABP, which would help understand the complex phenomena seen in experiments~\cite{xanthusagg1,xanthusagg2}.

\acknowledgements
U.B. acknowledges support from Science and Engineering Research Board (SERB), India
under Ramanujan Fellowship (Grant No. SB/S2/RJN-077/2018).

\end{document}


\title{Supplemental Material for ``Active Brownian Motion with Directional Reversals''}
\author{Ion Santra}
\affiliation{Raman Research Institute, Bengaluru 560080, India}
\author{Urna Basu}
\affiliation{Raman Research Institute, Bengaluru 560080, India}
\affiliation{S. N. Bose National Centre for Basic Sciences, Kolkata 700106, India}
\author{Sanjib Sabhapandit}
\affiliation{Raman Research Institute, Bengaluru 560080, India}
\maketitle


In this Supplemental Material we provide additional details of  calculations and numerical support for the results reported in the main text.

\section{Noise Correlation and Moments}

The DRABP dynamics follows the overdamped Langevin equations~(1) in the main text, 
\begin{subequations}
\bea
\dot x(t) &=& \zeta_x(t) \equiv v_0\sigma(t)\cos\theta(t),\\
\dot y(t) &=& \zeta_y(t) \equiv  v_0\sigma(t)\sin\theta(t), 
\eea
\label{eq:noise}
\end{subequations}
where $\sigma(t)=\pm 1$ is a dichotomous noise with reversal rate $\gamma$ and $\theta$ follows a Brownian motion with diffusion constant $D_R.$ To compute the moments of the position we need the auto-correlations of these effective noises. Since the $\sigma$ and $\theta$ processes are independent, it suffices to compute the correlations of $\sigma$ and $\cos\theta,\, \sin\theta$ separately.

We begin with the reversal process $\sigma(t)$. In this case, the propagator $\Psi(\sigma,t|\sigma_0,0),$ \ie, the probability that $\sigma(t)=\sigma$ given that 
$\sigma(0)=\sigma_0,$ is given by, 
\bea
\Psi(\sigma,t|\sigma_0,0)&=&\frac{1}{2}\Big(1+\sigma \sigma_0 e^{-2\gamma t} \Big).
\eea
The mean and the autocorrelation functions for the $\sigma$-process can be immediately obtained using the above equation,
\bea
\la \sigma (s)\ra &=&\sigma_0 e^{-2\gamma s}, \quad \text{and}~~
\la \sigma (s)\sigma (s')\ra = e^{-2\gamma |s-s'|}.
\label{sigma_corr}
\eea 
To obtain the autocorrelation functions for $\cos\theta$ and $\sin\theta$ we need the propagator for the $\theta$-process. Since $\cos \theta$ and $\sin \theta$ are periodic functions of $\theta$, the autocorrelations do not depend on the range of $\theta$, as long as it is an integral multiple of $2\pi$. For the sake of simplicity, we consider $ \theta \in(-\infty,  \infty)$, where the $\theta(t)$ distribution is a simple Gaussian at all times---the probability that $\theta(t)=\theta$ given that $\theta(0)=\theta_0$ is given by,
\bea
{\cal }P(\theta, t| \theta_0,0) = \frac 1{\sqrt{4 \pi \dr t}} \exp{\left[- \frac{(\theta -\theta_0)^2}{4 \dr t}\right]}.
\eea
Consequently,
\bea
\la \cos\theta(s)\ra &=&\cos\theta_0 e^{-D_R s},\quad \quad
\la \sin\theta(s)\ra =  \sin\theta_0 e^{-D_R s}, \label{eq:theta_av}
\eea
and the auto-correlation functions,
\begin{subequations}
\bea
\la \cos\theta(s)\cos\theta(s')\ra &=&\frac 12 \bigg[ e^{-D_R|s-s'|} + e^{-D_R(s+s'+2\,\text{min}[s,s'])}\cos2\theta_0 \bigg],\\
 \la \sin\theta(s) \sin\theta (s')\ra &=& \frac 12 \bigg[ e^{-D_R|s-s'|} -e^{-D_R(s+s'+2\,\text{min}[s,s'])}\cos2\theta_0 \bigg],\\
\la \cos\theta(s)\sin\theta(s') \ra &=& \frac{\sin ( 2\theta_0)}{2} e^{-D_R(s+s'+2\,\text{min}[s,s'])}.
 \label{theta_corr} 
\eea
\end{subequations}
Combining  Eqs.~\eqref{sigma_corr}-\eqref{theta_corr}, we have the mean of the effective noises,

\begin{subequations}
\bea
\la \zeta_x(s)\ra=v_0\la \sigma(s)\ra \la \cos\theta(s)\ra &=& v_0\sigma_0 \cos\theta_0 e^{-(D_R+2\gamma) s},\\
 \la \zeta_y(s)\ra=v_0\la \sigma(s)\ra \la \sin\theta(s)\ra &=& v_0 \sigma_0 \sin\theta_0 e^{-(D_R+2\gamma) s},\label{meannoise}
 \eea
\end{subequations}

and the auto-correlations,
\begin{subequations} 
 \bea
\la \zeta_x(s)\zeta_x(s')\ra=v_0^2\la \sigma(s)\sigma(s')\ra \la \cos\theta(s)\cos\theta(s')\ra &=&\frac {v_0^2}2 \bigg[ e^{-(D_R+2\gamma)|s-s'|} + e^{-2\gamma|s-s'|-D_R(s+s'+2\,\text{min}[s,s'])}\cos2\theta_0 \bigg], \\
\la \zeta_y(s)\zeta_y(s')\ra=v_0^2\la \sigma(s)\sigma(s')\ra \la \sin\theta(s)\sin\theta(s')\ra&=&\frac {v_0^2}2 \bigg[ e^{-(D_R+2\gamma)|s-s'|} - e^{-2\gamma|s-s'|-D_R(s+s'+2\,\text{min}[s,s'])}\cos2\theta_0 \bigg],\\
\la \zeta_x(s)\zeta_y(s')\ra=v_0^2\la \sigma(s)\sigma(s')\ra \la \cos\theta(s)\sin\theta(s')\ra&=&\frac {v_0^2 \sin 2\theta_0 }2\, e^{-2\gamma|s-s'|-D_R(s+s'+2\,\text{min}[s,s'])}.
\eea
\label{eff_noise_ac}
\end{subequations}
Note that from now on we always consider  $\sigma_0=1$ as that is the initial condition used in the main text.

\subsection{Position Moments} 
 
We compute the first two moments, namely, mean and variance of the position distribution of DRABP exactly using the effective noise auto-correlations,
\bea
\la x(t) \ra &=&  \int_0^t  ds~ \la \zeta_x(s) \ra , \quad  \la x^2(t)\ra =\int_0^t \int_0^t ds\, ds'~ \la \zeta_x(s)\zeta_x(s') \ra  \label{eq:x_moments}  \\
\la y(t) \ra &=&  \int_0^t  ds~ \la \zeta_x(s) \ra , \quad  \la x^2(t)\ra =\int_0^t \int_0^t ds\, ds'~ \la \zeta_x(s)\zeta_x(s') \ra.\label{eq:y_moments}
\eea

Using Eqs.~\eqref{meannoise}-\eqref{eff_noise_ac} we obtain the mean,
\bea
\la x(t) \ra &=&\frac{v_0 \cos\theta_0}{2\gamma +D_R}\left(1-e^{-t(2\gamma +D_R)}\right); \quad 
\la y(t) \ra =\frac{v_0 \sin\theta_0}{2\gamma +D_R}\left(1-e^{-t(2\gamma +D_R)}\right),
\label{mean_exact}
\eea
and the second moments,
\begin{subequations}
\bea
\la x^2(t)\ra &=&\frac{v_0^2 t}{(2\gamma+D_R)} + \frac{ v_0^2}{(2\gamma+D_R)^2} (e^{-(2\gamma+D_R)t}-1)+ \frac{v_0^2 \cos2\theta_0}{(3D_R-2 \gamma)} \left[\frac{e^{-4D_R t}-1}{4D_R}+\frac{1-e^{-(D_R+2\gamma)t}}{(D_R+2\gamma)}\right],\\
\la y^2(t)\ra &=& \frac{v_0^2 t}{(2\gamma+D_R)} + \frac{ v_0^2}{(2\gamma+D_R)^2} (e^{-(2\gamma+D_R)t}-1)- \frac{v_0^2 \cos2\theta_0}{(3D_R-2 \gamma)} \left[\frac{e^{-4D_R t}-1}{4D_R}+\frac{1-e^{-(D_R+2\gamma)t}}{(D_R+2\gamma)}\right].~~~~
\eea
\label{2ndmoment_exact}
\end{subequations}

The presence of the two time-scales $D_R^{-1}$ and $\gamma^{-1}$ gives rise to four distinct dynamical regimes characterized by different dynamical behaviors: a short-time regime (I) $t\ll \min (\gamma^{-1}, D_R^{-1})$, two intermediate-time regimes 
(II) $\gamma^{-1}\ll t\ll D_R^{-1}$ (accessible for $\gamma > \dr$) and (III) $\dr^{-1}\ll t \ll \gamma^{-1}$ (accessible for $\gamma < \dr$), and the long-time regime (IV) $t\gg \max (\gamma^{-1}$, $D_R^{-1}).$

In the following we look at how the variances $\la x^2(t)\ra_c = \la x^2(t)\ra - \la x(t)\ra^2$ and  $\la y^2(t)\ra_c =\la y^2(t)\ra - \la y(t)\ra^2$ behave in these regimes.

\subsubsection{Short-time regime (I)}
The behavior of the variance in this regime, \ie, for $t \ll \min(\gamma^{-1}, D_R^{-1}),$  can be obtained by simply expanding Eqs.~\eqref{2ndmoment_exact} in a Taylor series in $t,$
\begin{subequations}
\bea
\la x^2(t)\ra_c  &=& \frac{v_0^2\,t^3}{3}\left(D_R+2\gamma-(D_R-2\gamma)\cos2\theta_0\right)+\text{O}(t^4),\\
 \la y^2(t)\ra_c  &=& \frac{v_0^2\,t^3}{3}\left(D_R+2\gamma+(D_R-2\gamma)\cos2\theta_0\right)+\text{O}(t^4).
\eea
\end{subequations}
Clearly, there is an anisotropy in the system if we begin from arbitrary $\theta_0$ (except when $\cos 2\theta_0=0$), on the other hand, if the initial orientation is chosen uniformly in $[0,2\pi]$, then $\la x^2(t)\ra_c=\la y^2(t)\ra_c=\frac{v_0^2t^3}{3}(\dr+2\gamma) + O(t^4)$. 

\subsubsection{Intermediate regime (II)}
 In this regime, $\gamma t \gg 1$ while $\dr t \ll 1,$ hence,  the behavior of the variance is obtained by neglecting terms $\sim e^{-2\gamma t}$ and then expanding the resulting expression in a series in $D_R t,$
\begin{subequations} 
 \bea
 \la x^2(t)\ra_c\approx \frac{v_0^2t}{2\gamma}(1+\cos2\theta_0)-\frac{v_0^2t^2 D_R}{\gamma}\cos2\theta_0 +\text{O}(t^3),\\
 \la y^2(t)\ra_c\approx \frac{v_0^2t}{2\gamma}(1-\cos2\theta_0)+\frac{v_0^2t^2 D_R}{\gamma}\cos2\theta_0 +\text{O}(t^3).
 \eea
 \label{var_int}
 \end{subequations}
 Thus in this regime, the anisotropy persists. This is evident from the variances along and orthogonal to the initial orientation $\theta_0$. Setting $\theta_0=0$ in Eq.~\eqref{var_int} yields, $\la x^2(t)\ra_c\equiv\la x_{\parallel}^2(t)\ra_c\propto t$ and $\la y^2(t)\ra_c\equiv\la x_{\perp}^2(t)\ra_c\propto t^2$.
 
 \subsubsection{Intermediate regime (III)}
 
In this regime $\dr t \gg 1$ while $\gamma t \ll 1$ and one can neglect terms $\sim e^{-\dr t}.$ Expanding the resulting expression in a series of $\gamma t$ we get the variance behavior, In the regime (III), \ie, for $ D_R^{-1} \ll t \ll \gamma^{-1},$ we get,
 \bea
 \la x^2(t)\ra_c=\la y^2(t)\ra_c\approx \frac{v_0^2 t}{D_R}
 \eea
 which indicates that if $D_R>\gamma$, the anisotropy vanishes already in the intermediate regime and the motion becomes diffusive with an effective diffusion coefficient $v_0^2/(2D_R).$ 
 
 \subsubsection{Long-time regime (IV)}
  Finally, for $t \gg \max(\gamma^{-1},D_R^{-1}),$ we have,
 \bea
 \la x^2(t)\ra_c = \la y^2(t)\ra_c \approx 2D_{\text{eff}} t.
 \eea
 The dynamics is isotropic and diffusive with an effective diffusion constant $D_{\text{eff}}=\frac{v_0^2}{2(D_R+2\gamma)}$. Note that this $D_{\text{eff}}$ is the same as obtained in \cite{grossman1} for the case when the internal clock has a single state.

 \section{Effective dynamics in the different regimes}\label{sec:eff_dynamics} 
 
The DRABP dynamics in the four different temporal regimes can be effectively described by some simpler versions of Eq.~\eqref{eq:noise}. In this section we summarize these effective dynamics in the different regimes.

\subsubsection{Short-time regime (I)} 
 In this regime $t$ is much smaller than both time-scales $\dr^{-1}$ and $\gamma^{-1}.$ Let us suppose that the particle starts from an initial orientation $\theta_0$, then the effective noises in Eq.~\eqref{eq:noise} can be written as,
 \begin{subequations}
 \bea
\zeta_x (t) &=& v_0 \sigma(t) (\cos \theta_0 \cos \phi(t) - \sin \theta_0 \sin \phi(t))\\
\zeta_y (t) &=& v_0 \sigma(t) (\sin \theta_0 \cos \phi(t)+\cos \theta_0 \sin \phi(t)) 
\eea
\label{eq:noise2}
\end{subequations}
where $\phi(t) = \sqrt{2 \dr} \int_0^t ds~ \eta(s)$ is a standard Brownian motion.  At times $t\ll D_R^{-1}$,  $\phi(t)~\sim\sqrt{D_R t} \ll 1$, so we can use the approximation $\cos \phi \simeq 1$ and $\sin \phi \simeq \phi$ to the leading order in $\phi.$ Equations~\eqref{eq:noise2} then reduce to,
\begin{subequations}
\bea
\zeta_x(t)&\approx & \sigma(t)(A-B\phi(t))\\
\zeta_y(t)&\approx & \sigma(t)(B+A\phi(t)). 
\eea
\label{eq:short_langevin}
\end{subequations}
where we have used $A=v_0\cos\theta_0$ and $B=v_0\sin\theta_0$ for notational simplicity. We use this form of the effective noise correlation to compute the position distribution in this regime in Sec.~\ref{sec:short_t}.

\subsubsection{Intermediate-time regime (II)}

 In this regime $\phi$ does not change appreciably, since $D_R t  \ll 1$, as in regime (I). So the approximations $\cos \phi \simeq 1$ and $\sin \phi \simeq \phi$ remain still valid. On the other hand there is a large number of directional reversals as $\gamma t\gg 1$. Thus, in this regime, the effective noise can be approximated as,
\begin{subequations}
\bea
\zeta_{x}(t)&=&\xi(t)(A-B\phi(t)), \\
\zeta_{y}(t)&=&\xi(t)(B+A\phi(t)),
\eea
\label{langevin_intermediate}
\end{subequations}
where $\xi(t)$ is a Gaussian white noise with the following properties,
\bea
\la \xi(t)\ra &=&0, \quad
\la \xi(t)\xi(t')\ra = \frac{1}{\gamma}\delta(t-t').
\label{intermed_corr}
\eea
We use this approximation of the effective noise to calculate the position distribution (Sec.~\ref{sec:int_pos}) and persistence exponent (Sec.~\ref{sec:int_surv}) of the DRABP in regime (II).

\subsubsection{Intermediate-time regime (III) and long-time regime (IV)}
These regimes can be accessed using $\dr\to\infty$. For large $\dr$,  effective noise correlations Eqs.~\eqref{eff_noise_ac} reduce to, 
 \bea
 \la \zeta_{a}(s)\zeta_{b}(s')\ra &\approx &\delta_{a,b}\frac {v_0^2}2 \exp\left(-(D_R+2\gamma)|s-s'|\right). \label{eff_corr_lt}
 \eea
In regime (IV), taking the limits $D_R\to\infty$ and $\gamma\to \infty$, while keeping $\dr/\gamma$ and $2D_{\text{eff}}=\frac{v_0^2}{(D_R+2\gamma)}$ finite, we get,
 \bea
 \la \zeta_{a}(s)\zeta_{b}(s')\ra &=& 2D_{\text{eff}}\delta_{a,b}\delta(s-s').
 \eea 
 This is the effective noise correlation used in regime (IV) in the main-text. On the other hand, in the limit $D_R\to\infty$ and $\gamma/\dr\ll 1 $, with $v_0^2/\dr$ finite, we get,
 \bea
 \la \zeta_{a}(s)\zeta_{b}(s')\ra &=& \delta_{a,b}\frac{v_0^2}{D_R}\delta(s-s'),
 \eea  
 which is used to calculate the position distribution of the DRABP in regime (III) in the main-text.

\section{Position distribution: Short-time regime}\label{sec:short_t}
In this section we provide the details of the computation leading to the marginal position distribution in the short-time regime (I) $t \ll \min(\gamma^{-1},\dr^{-1})$,  \ie, when the time is much smaller than both the characteristic time scales of the system. The dynamics in this regime is governed by the Langevin equations $\dot{x}=\zeta_x(t)$ and $\dot{y}=\zeta_y(t)$ where the effective noise $\zeta_x(t)$ and $\zeta_y(t)$ are given by Eq.~\eqref{eq:short_langevin}.

\begin{figure}[h]
\centering\includegraphics[width= 9 cm]{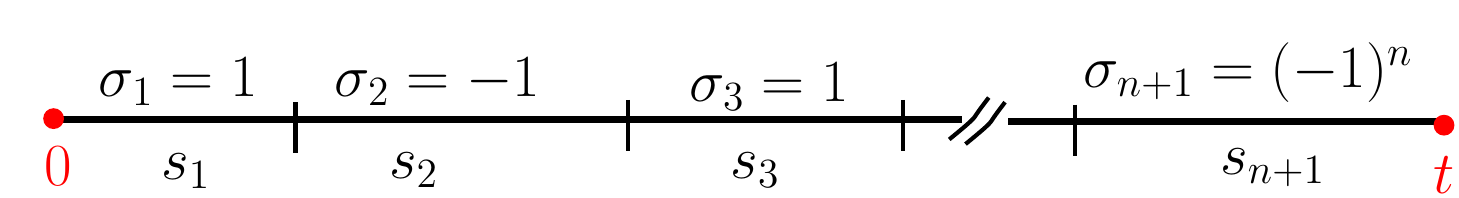}
\caption{Schematic representation of the reversal process: $s_i$ denotes the interval between $i^{th}$ and $(i+1)^{th}$ reversal events during which $\sigma_i = (-1)^{i-1}$ remains constant.}
\label{interval}
\end{figure}

Now, let us assume that during time $t$ there are $n$ orientational reversals. We can thus divide the duration $[0,t]$ into $n + 1$ intervals, such that $\sigma$ changes  sign at the beginning of each interval and remains constant throughout the interval. Let $s_i$ be the duration of the $i$-th interval as shown in Fig.~\ref{interval} and $\sigma_i = (-1)^{i-1}$ denotes the value of $\sigma$ in this interval. For the sake of convenience we also define $t_i = \sum_{j=1}^i s_j$ which is the total time elapsed before the start of the $(i+1)$-th interval. Obviously, $t_0=0$ and $t_{n+1} = t.$ 

For a given trajectory $\{ \sigma_i, s_i \},$ the final position of the particle can then be expressed as,
\begin{subequations}
\bea
x(t) &=& A \sum_{i=1}^{n+1}{\sigma_i s_i} - B \sum_{i=1}^{n+1} \sigma_i z_i, \\
y(t) &=& B \sum_{i=1}^{n+1}{\sigma_i s_i} + A \sum_{i=1}^{n+1} \sigma_i z_i \eea
\label{eq:xy_sigma}
\end{subequations}
where we have denoted $z_i = \int_{t_{i-1}}^{{t_i}} ds~\phi(s).$ 
Since $\phi(s)$ is an ordinary Brownian motion, its integral should follow a Gaussian distribution--- in fact, $\{ z_i; i=1,2 \cdots n+1 \}$ form a set of $(n+1)$ correlated Gaussian variables with the correlation matrix $C_{ij} = \la z_i z_j \ra.$ The linear combination $\sum_{i=1}^{n+1} \sigma_i z_i$ then also follows a Gaussian distribution with the variance 
\bea
b_n = \sum_{i,j=1}^{n+1} \sigma _i \sigma_j C_{ij}. \label{eq:Z-var}
\eea

From Eq.~\eqref{eq:xy_sigma}, we can then write the marginal position distributions for a given trajectory $\{ \sigma_i, \tau_i \},$
\begin{subequations}
\bea
{\cal P}(x,\{ \sigma_i, s_i \}) &=& \frac 1{B \sqrt{2 \pi b_n}} \exp{\left[- \frac{(x- A\sum_{i=1}^{n+1} \sigma_i s_i)^2}{ 2 b_n B^2 } \right]}, \\
{\cal P}(y,\{ \sigma_i, s_i \}) &=& \frac 1{A \sqrt{2 \pi b_n}} \exp{\left[- \frac{(y- B\sum_{i=1}^{n+1} \sigma_i s_i)^2}{ 2 b_n A^2 } \right]}. 
\eea
\label{eq:Pxy_sigma}
\end{subequations}
Note that, for notational simplicity we have used the same letter ${\cal P}$ to denote both $x$ and $y$ distributions. 
The variance $b_n$ is obtained from Eq.~\eqref{eq:Z-var} using the correlation matrix $C_{ij}$ which can be computed explicitly using the auto-correlation of the Brownian motion $\la \phi(s) \phi(s') \ra = 2 D_R \min(s,s'),$
\bea
C_{ij} = \left \{ \begin{split}
D_R  (t_i^2-t_{i-1}^2)(t_j-t_{j-1}) & ~\text{for} ~~ i < j \cr
D_R  (t_j^2-t_{j-1}^2)(t_i-t_{i-1}) & ~\text{for} ~~j < i \cr
\frac {2D_R}3 (t_i-t_{i-1})^2(t_i+2 t_{i-1}) & ~\text{for} ~~ i=j.
\end{split} \label{eq:Cij}
\right.
\eea

To obtain the actual position distribution $P(x,t)$\footnote{We use the same notation $P(\cdot)$ to denote all probability density functions. The number of arguments as well as the actual functional form is different depending on the context.} we need to take into account the contributions from all possible trajectories $\{\sigma_i, s_i;i=1,2,\dotsc ,n+1\}$ with all possible values of $n$,
\bea
P(x,t) = \sum_{n=0}^\infty \gamma^n e^{-\gamma t} P_n(x,t)
\label{shorttimex}
\eea
where $P_n(x,t)$ denotes the contribution from the trajectories with $n$ number of reversals,
\bea
P_n(x,t) &=& \int_{0}^t \prod_{i=1}^{n+1} d s_i~ \delta(t - \sum_{i}s_i) {\cal P}(x,\{ \sigma_i, s_i \}). \label{pnx}
\eea 
This result is quoted as Eq.~(3) in the main text.
Similarly, for the $y$-component,
\bea
P(y,t) &=& \sum_{n=0}^\infty \gamma^n e^{-\gamma t} P_n(y,t) = \sum_{n=0}^\infty \gamma^n e^{-\gamma t} \int_{0}^t \prod_{i=1}^{n+1} d s_i~ \delta(t - \sum_{i}s_i)~{\cal  P}(y,\{ \sigma_i, s_i \}).\label{py}
\eea

As discussed in the main text, for $t \ll \gamma^{-1},$ it suffices to look only at the first few terms as the average number of reversals is given by $\la n \ra =\gamma t$ during a time-interval $t.$  In fact, this can be thought of as a perturbative approach in $\gamma,$ which should work well for the DRABP in the regime $t \ll \min(\gamma^{-1}, D_R^{-1}).$ In the following we  compute the first few terms explicitly and illustrate, by comparing the corresponding analytical prediction with the data from numerical simulations, that the series converges reasonably fast.  We restrict ourselves to marginal $x$-distribution only, the $y$-distribution can be obtained following the same procedure.

\begin{figure}[h]
\includegraphics[width=8.8cm]{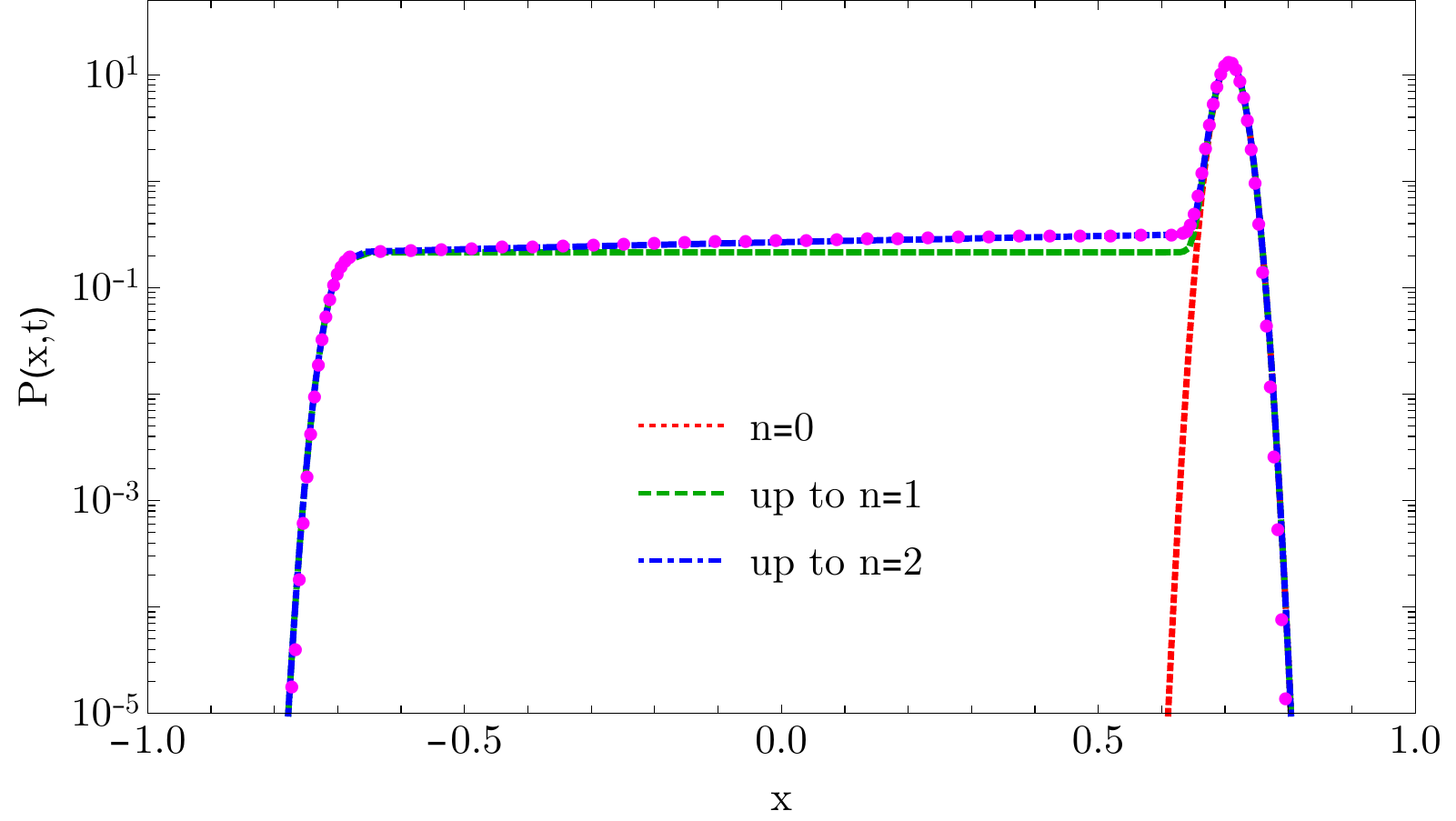}
\caption{Convergence of the infinite series \eqref{shorttimex}: Plot of $P(x,t=1)$ for $\gamma=0.5$, $D_R=10^{-3}$ and $\theta_0=\pi/4$. The magenta symbols represent the result obtained from simulations, while the dashed lines indicate $P(x,t)$ in Eq.~\eqref{shorttimex} calculated upto specified terms as mentioned in the legends.}\label{f:convergence} 
\end{figure}

For $n=0,$ there are no reversal events. In this case $b_0=2 D_R t^3/3$  and the corresponding contribution is a simple Gaussian,
\bea
P_0(x,t) = \frac{\sqrt{3}}{2\sqrt{\pi D_R}B t^{3/2}}\exp\left(-\frac{3(x-A t)^2}{4B^2 D_R t^3}\right).
\eea
For $n=1,$ the velocity reverses once, after some time $s_1 \in [0,t].$ In this case, we have, from Eqs.~\eqref{eq:Z-var} and \eqref{eq:Cij},
\bea
b_1 = \frac {2 D_R}3 [(s_1+s_2)^3- 6 s_1^2 s_2].
\eea
Corresponding contribution to the position distribution is given by,
\bea
P_1(x,t) = \sqrt{\frac 3{4\pi D_R B^2 t^3}}\int_{0}^{t} ds_1~ \frac{\exp{\left(-\frac{3(x-A(2s_1-t))^2}{4B^2D_R(t^3 - 6 s_1^2 (t- s_1))}\right)}}{\sqrt{t^3 - 6 s_1^2( t -s_1)}}.\n
\eea 
Similarly, for $n=2,$ \ie, two reversals, 
\bea
b_2& =&  \frac{2 D_R}{3} \Bigg[\left(s_1+s_2+s_3\right)^3-6 s_2 \left(s_1+\tau _2\right) \left(s_1+s_2+s_3\right) +6 s_2^2 \left(2 s_1+s_2\right)\Bigg],
\eea 
and 
\bea
P_2(x,t) &=& \sqrt{\frac 3{4\pi D_R B^2 t^3}} \int_{0}^{t} ds_1\int_{0}^{t-s_1}ds_2\frac{\exp \left(\frac{-3 (y-B (t-2 s_2))^2}{4 A^2 D \left(s_1^3+(t-s_1-s_2)^3+s_2^3\right)}\right)
}{\sqrt{s_1^3+(t-s_1-s_2)^3+s_2^3}},
\eea
and so on. Clearly, $P_n(x,t)$ can be obtained  systematically by evaluating the integrals numerically.

To illustrate the convergence of $P(x,t)$ in Eq.~\eqref{shorttimex}, in Fig.~\ref{f:convergence} we plot the contributions from the first few terms separately for $\gamma=0.5$ and $t=1$ (the uppermost curve in Fig.~2(a) in the main text). Clearly, the effect of reversal is immediately visible from the $n=1$ term, which changes the shape of the distribution drastically, adding a plateau around the origin, in addition to the Gaussian peak near $x \sim v_0 t \cos \theta_0$; the higher order terms only add quantitative corrections. As expected, when $\gamma$ is increased, we obtain better quantitative match by including higher order terms (see Fig.~\ref{f:convergence}).

We conclude the discussion about the short-time regime with a brief comment.
If the particle starts from a uniform initial orientation in $[0,2\pi]$, then the  resulting distribution is always isotropic. The corresponding marginal distribution can be obtained by integrating Eq.~\eqref{shorttimex} over $\theta_0$,
\bea
\bar{P}(x,t)=\frac{1}{2\pi}\int_0^{2\pi} d\theta_0~ P(x,t).
\label{shortt_uniform}
\eea

\begin{figure}[t]
\includegraphics[width= 8 cm]{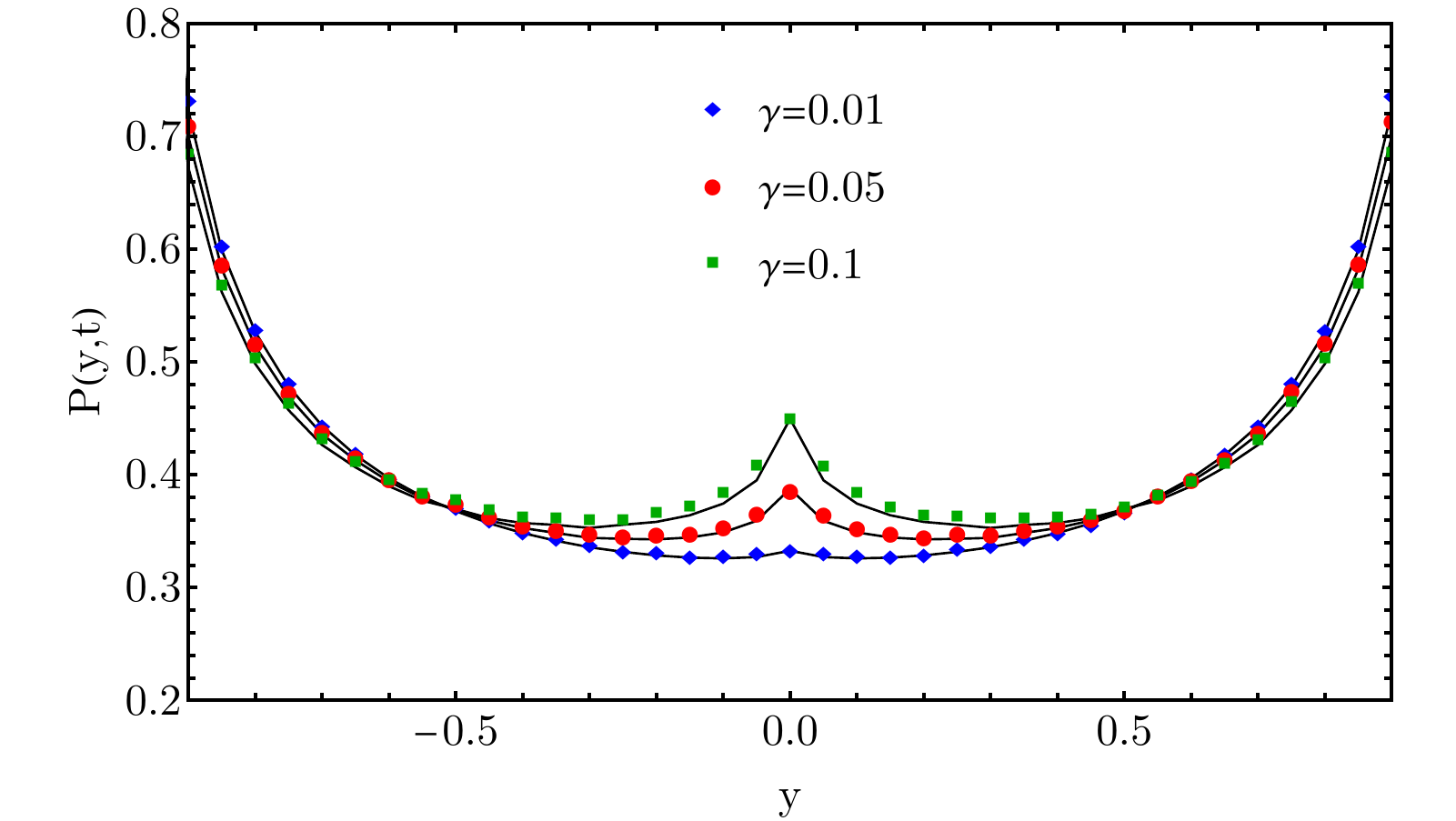}
\includegraphics[width= 7 cm]{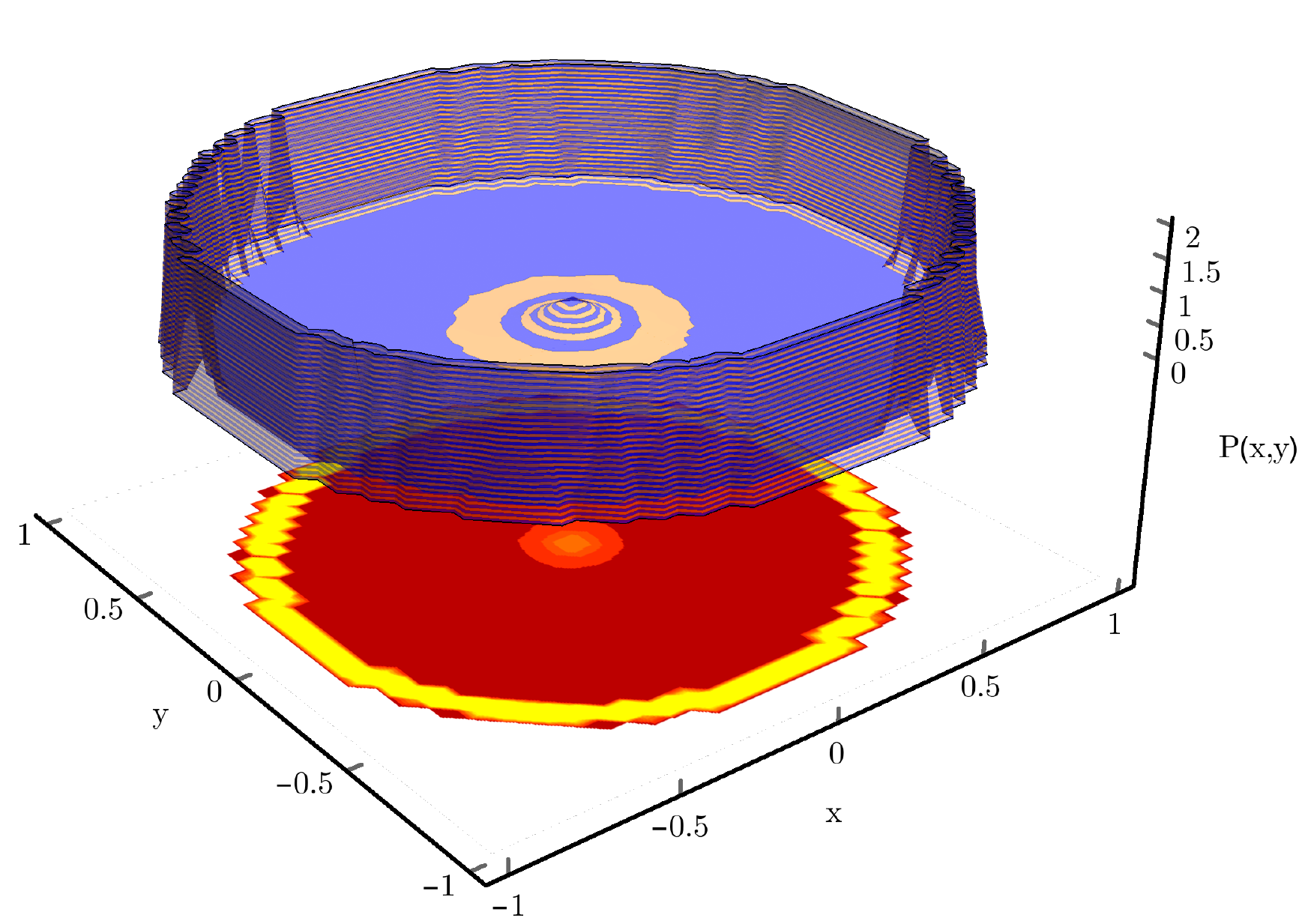}
\caption{Left: Comparison of probability distribution for uniform initial orientation obtained from numerical simulation and calculating upto $n=2$ term of the series in Eq.~\eqref{shortt_uniform}. Right: $P(x,y,t)$ and the corresponding contour plot for $t=1$ and $\gamma=0.1,$ $D_R=0.01$ obtained from numerical simulation of Eq.~(1) with uniform initial orientation}.
\label{uniformdist}
\end{figure}

There is a central peak in this case which increases with the increase in $\gamma$. Using Eq.~\eqref{shortt_uniform} and the first few terms of Eq.~\eqref{shorttimex} one can get a good approximation for this distribution as shown in Fig.~\ref{uniformdist}.

\section{Position distribution: Intermediate-time regime (II)}\label{sec:int_pos}

In this section we focus on the position distribution in the regime II  and provide a detailed derivation of the Eqs.~(2) and (10) in the main text.  As discussed in the main text and Sec.~\ref{sec:eff_dynamics}, the effective noises governing the DRABP dynamics in this regime, \ie, for $\gamma^{-1} \ll t \ll D_R^{-1}$ can be approximated by Eq.~\eqref{langevin_intermediate},
We can thus write the characteristic function for the joint distribution as,
\bea
\la \exp{(i\bm{k}\cdot \bm{x})}\ra &=&\left\la \exp{\left(i\int_0^t ds\,\xi(s)\left[k_x\left(A-B\phi(s)\right)+k_y\left(B+A\phi(s)\right)\right]\right)} \right\ra_{(\xi, \phi)},
\eea 
with $A=v_0\cos\theta_0,\,B=v_0\sin\theta_0$. Here the averaging is over both $\{\xi(t)\}$ and $\{\phi(t)\}$ trajectories and $\bm k=\begin{pmatrix}
k_x\\
k_y
\end{pmatrix}$, $\bm x=\begin{pmatrix}
x\\
y
\end{pmatrix}$. Now, for a given trajectory $\{\phi(t)\}$, the  averaging over $\{\xi(t)\}$ can be done immediately to yield,
\bea
\la \exp{(i\bm{k}\cdot \bm{x})}\ra &=&\left\la \exp\left[-\frac{1}{2} \bm{k}^\text{T} \bm\Sigma(t) \bm{k} \right] \right\ra_{\phi},
\label{charf}
\eea
where the average is now over the $\phi$-process only. The covariance matrix  is given by,
\bea
\bm \Sigma(t)= \begin{bmatrix}
 \la x^2(t)\ra_\xi && \la x(t)y(t)\ra_\xi\\
 \la x(t)y(t)\ra_\xi && \la y^2(t)\ra_\xi
\end{bmatrix}=  \frac{1}{\gamma}\begin{bmatrix}
\int_{0}^t ds~ (A-B\phi(s))^2 && \int_{0}^t ds~ (A-B\phi(s))(B+A \phi(s))\\
\\
\int_{0}^t ds~ (A-B\phi(s))(B+A \phi(s)) && \int_{0}^t ds~ (B+A\phi(s))^2
\end{bmatrix}.
\eea
 Remembering that $\phi (s)$ is a standard Brownian motion, the rhs of Eq.~\eqref{charf} can be evaluated using  path integral~\cite{brownianfunctionals},
\bea
\la \exp{(i\boldsymbol{k} \cdot \boldsymbol{x})}\ra &=&\int_{-\infty}^{\infty} d X \int_{0}^{X} {\cal D}\phi\, \exp\left[-\int_0^t ds\left(\frac{\dot{\phi}^2}{4D_R}  -\frac{1}{2\gamma}\left(k_x(A-B\phi)+k_y(B+A\phi)\right)^2\right)\right]\cr
&=& \int_{-\infty}^{\infty} d X \int_{0}^{X} {\cal D}\phi\, \exp\left[-\int_0^t ds\left( \frac{Z_1^2}{2\gamma}(\phi+Z_2/Z_1)^2+\frac{\dot{\phi}^2}{4D_R}\right)\right],\label{eq:genfn}
\eea
where $Z_1=(k_yA-k_xB)$ and $Z_2=(k_xA+k_yB)$. Using the variable shift $\phi \to \phi+\frac{Z_2}{Z_1}$ and $X \to X+\frac{Z_2}{Z_1}$, Eq.~\eqref{eq:genfn} reduces to,
\bea
\la \exp{(i\boldsymbol{k}.\boldsymbol{x})}\ra &=&\int_{-\infty}^{\infty} dX \int_{Z_2/Z_1}^{X}{\cal D}\phi~ \exp\left[-\int_0^t ds\left(\frac{\dot{\phi}^2}{4 D_R}+\frac{Z_1^2}{2\gamma}\phi^2\right)\right].
\eea
The form of the path integral in the above equation corresponds to the imaginary time propagator of a quantum harmonic oscillator with Hamiltonian $H=-\frac{\hbar^2}{2m}\frac{d^2}{d x^2}+\frac{1}{2}m\omega^2 x^2$, upon setting $\hbar=1$, $m=\frac{1}{2D_R}$ and $\omega^2=\frac{2Z_1^2 D_R}{\gamma}$. It propagates from initial position $\frac{Z_2}{Z_1}$ to the final position $X$ in time $t$. Thus, we have,
\bea
\la \exp{(i\boldsymbol{k}.\boldsymbol{x})}\ra &=&\int_{-\infty}^{\infty} dX\, U(X,Z_2/Z_1,t), \label{eq:gen_U}
\eea
where, $U(X_f,X_i,t)$ is the propagator of a quantum harmonic oscillator with initial and final points $X_i$ and $X_f$ respectively in imaginary time $t$. This is well known in literature \cite{feynmann} and with the mappings mentioned earlier we have,
\bea
 U(X_f,X_i,t)=\sqrt{\frac{\omega}{4\pi D_R \sinh(\omega t)}}\exp\left[-\frac{\omega}{4D_R\sinh(\omega t)}\bigg(\left(X_f^2+X_i^2\right)\cosh(\omega t) -2X_f X_i\bigg) \right].
\eea
Using the above expression in Eq.~\eqref{eq:gen_U} and performing the integral over $X$, we obtain,
\bea
\la \exp{(i\boldsymbol{k}.\boldsymbol{x})}\ra &=&\frac{1}{\sqrt{\cosh\omega t}}\exp\left[-\frac{\omega Z_2^2 \tanh\omega t}{4D_R Z_1^2}\right].
\eea
Substituting $Z_1=(k_yA-k_xB)$ and $Z_2=(k_xA+k_yB)$ in the above equation, we get Eq.~(10) in the main text.
The characteristic functions for $x$ and $y$ marginal distributions are obtained by taking $k_y=0$ and $k_x=0$ respectively, 
\begin{subequations}
\bea
\la e^{ikx}\ra &=\frac{1}{\sqrt{\cosh\omega_x t}}\exp{\left[-\frac{\omega_x}{4D_R} \cot^2 \theta_0 \tanh \omega_x t \right]}, \\
\la e^{iky}\ra &=\frac{1}{\sqrt{\cosh\omega_y t}}\exp{\left[-\frac{\omega_y}{4D_R} \tan^2 \theta_0 \tanh \omega_y t \right]}.
\eea
\label{marginal_char}
\end{subequations}
where $\omega_x= k B\sqrt{\frac{2D_R}{\gamma}}  $ and $\omega_y= k A \sqrt{\frac{2D_R}{\gamma}}.$ As mentioned in the main text, we are interested in the directions parallel and perpendicular to the initial orientation, denoted by $x_{\parallel}$ and $x_{\perp}$ respectively. This can be obtained by setting $\theta_0=0$, in which case $x_{\parallel}\equiv x$ and $x_{\perp}\equiv y$. From Eq.~\eqref{marginal_char}, we get 
\bea
\la e^{ikx_{\perp}}\ra=\left[\cosh \left(v_0kt\sqrt{\frac{2D_R}{\gamma}} \right)\right]^{-1/2} \qquad \text{and}\qquad \la e^{ikx_{\parallel}}\ra = \exp\left(-\frac{k^2v_0^2 t}{2\gamma}\right).
\eea
This can be inverted exactly to yield,
\bea
P(x_\perp,t) =   \frac {1}{v_0t}\sqrt{\frac \gamma{8 \dr}} f\left(\frac {x_\perp}{v_0t}\sqrt{\frac \gamma{8 \dr}}\right),\qquad\text{and}\qquad
P(x_{\parallel},t)&=&\frac{\sqrt{\gamma}}{v_0\sqrt{ t}}h\left(\frac{x_{\parallel}\sqrt{\gamma}}{v_0\sqrt{t}}\right),
\eea
with
\bea
f(z) =\frac 1 {\sqrt {2 \pi^3}} \Gamma\left(\frac 14 + i z \right) \Gamma\left(\frac 14 - i z\right),\qquad \text{and}\qquad h(z)=\frac{1}{\sqrt{2\pi}}\exp(-z^2/2). \label{intermediate_exact}
\eea
%
%
where $\Gamma(z)$ denotes the gamma function. The scaling form for $x_{\perp}$ is quoted in Eq.~(2) in the main text.

\begin{figure}[h]
\includegraphics[width=11 cm]{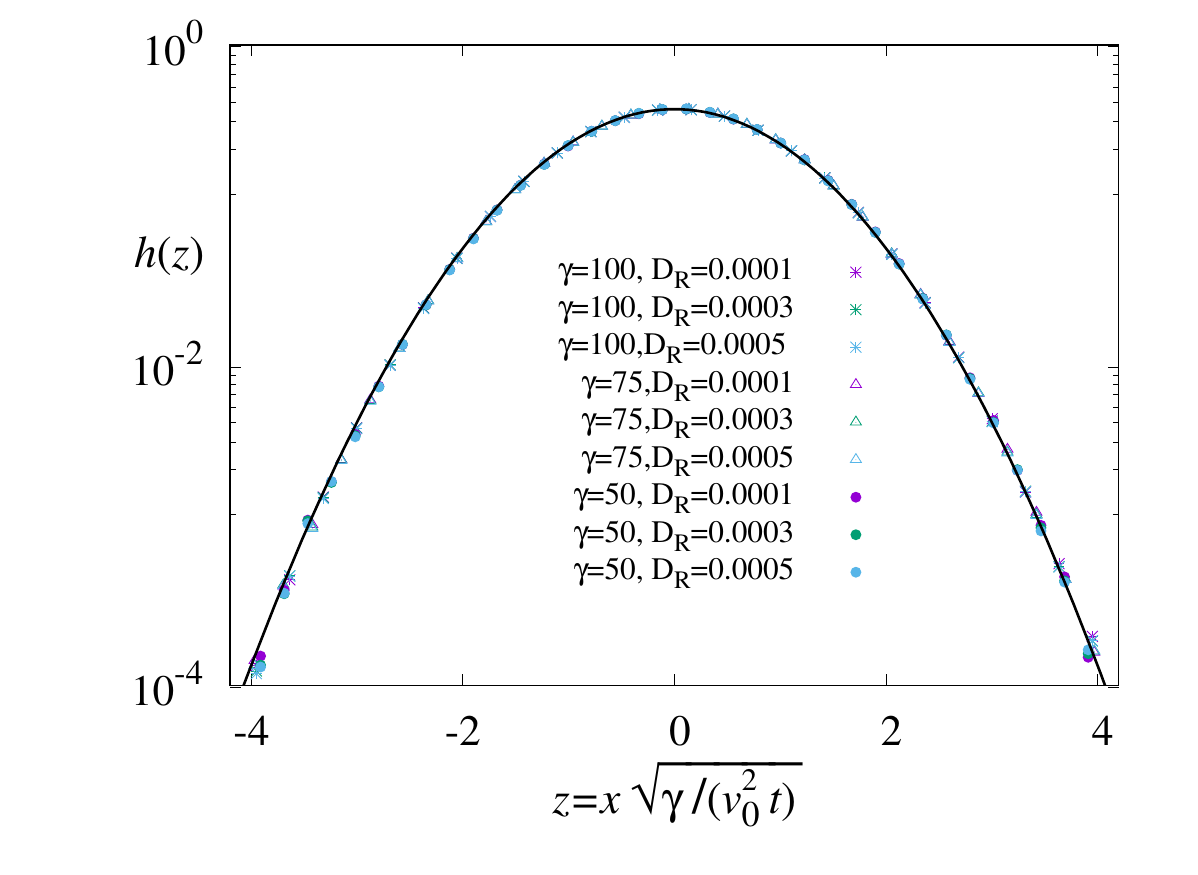}
\caption{Plot for distribution of the scaled variable $z=x\sqrt{\frac{\gamma}{v_0^2 t}}$ for $\theta_0=0$ (for which $x=x_{\parallel}$) in the intermediate regime (II) for $t=10.$ The solid black line denotes the scaling function $h(z)$ in Eq.~\eqref{intermediate_exact}. Note that the distribution of $x_{\parallel}$ is independent of $D_R$ in this regime.}
\label{f:parallel}
\end{figure}

 Clearly, while $x_{\parallel}$ follows a  diffusive scaling with a Gaussian distribution, $x_{\perp}$ is described by a non-trivial scaling function with ballistic scaling. Figure~\ref{f:parallel} compares the scaling function $h(z)$ in Eq.~\eqref{intermediate_exact} with numerical simulations.

\section{Survival Probability along the orthogonal direction in Regime II}\label{sec:int_surv}
In this section we work out in detail the survival probability of a DRABP in the intermediate time regime ($\gamma^{-1}\ll t\ll D_R^{-1}$) along the direction orthogonal to the initial orientation. We obtain this by setting $\theta_0=0$ where $x_{\perp}\equiv y$.
Let $S_y(t;y_0)$ denote the probability that a particle starting from $(0,y_0)$ with an initial orientation $\theta_0=0$ has not crossed the $y=0$ line till time $t$. Mathematically,
\bea
S_y(t;y_0) = \int_0^\infty \id y ~ P(y,t;y_0) \label{eq:SP}
\eea
where $P(y,t;y_0)$ is the marginal probability distribution in the presence of an absorbing wall at $y=0$, starting from the initial position $y_0.$  The survival probability in this regime is actually determined by trajectories which have already survived regime (I). Thus, in principle, one should take into account dynamics of both the regimes (I) and (II). However, the regime (I) almost vanishes for $\gamma\gg 1$ and it suffices to consider the effective dynamics in regime (II) only. Hence, we start with the 
Langevin equation along the $y$-direction [Eq.~\eqref{langevin_intermediate} with $\theta_0=0$] in the intermediate regime (II),
\bea
\dot{y}=v_0\,\xi(t)\phi(t),
\label{langevin_surv}
\eea
where $\phi(t)$ is a Brownian motion. 
We can write the corresponding forward Fokker-Planck (FP) equation for $P(y,\phi,t),$ \ie, the probability that  $y(t)=y$ and $\phi(t)=\phi,$ 
\bea
\frac{\partial }{\partial t} P(y,\phi,t) &=&\frac{v_0^2\phi^2}{2\gamma}\frac{\partial^2 }{\partial y^2} P(y,\phi,t) +D_R\frac{\partial^2 }{\partial \phi^2} P(y,\phi,t).
\label{fp_intermed}
\eea
Note that, for notational simplicity we have suppressed the initial position dependence. 
We need to solve this FP equation with the initial condition $P(y,\phi,0)=\delta(y-y_0)\delta(\phi)$ and boundary conditions $P(y,\phi,t)\to 0$ as $\phi(t)\to\pm \infty$ and $P(0,\phi,t)=P(\infty,\phi,t)=0.$ For simplicity, we make a change of variable $\frac {v_0^2t}{2\gamma} \to t$ and define  $\Lambda^2=2\gamma D_R/v_0^2 $.
 Equation~\eqref{fp_intermed} then becomes
\bea
\frac{\partial }{\partial t} P(y,\phi,t)&=&\phi^2 \frac{\partial^2 }{\partial y^2} P(y,\phi,t)+\Lambda^2\frac{\partial^2 }{\partial \phi^2} P(y,\phi,t).
\label{fp_i}
\eea
The absorbing boundary condition at $y=0$ can be taken care of by using the  $\sin$-eigenbasis $\sin (ky)$ with $k \ge 0.$  It is also convenient to take a Laplace transform w.r.t. time $t$,
\bea
\tilde P(k,\phi,s) = \int_0^\infty dt~ e^{-st} \int_0^\infty dy~ \sin(k y) \, P(y,\phi,t).
\eea
Equation~\eqref{fp_i} reduces to an ordinary second order differential equation in terms of $\tilde P(k,\phi,s)$,
\bea
\Lambda^2\frac{\id^2 }{\id \phi^2} \tilde{P}(k,\phi,s)-(s+ \phi^2 k^2) \tilde{P}(k,\phi,s)= - \sin(ky_0)\delta(\phi).\label{eq:Ptdiff}
\eea
with the boundary condition $\tilde P(k,\phi,s) \to 0$ for $\phi \to \pm \infty.$ For $\phi\ne 0,$ The general solution of Eq.~\eqref{eq:Ptdiff} is given by
\bea
\tilde{P}(k,\phi,s)=a~ D_{-q}\left(\phi \sqrt{\frac{2k}\Lambda}\right)+b~ D_{-q}\left(-\phi \sqrt{\frac{2k}\Lambda}\right),
\eea
where $q=\frac 12 (1 +\frac{s}{k\Lambda})$, $D_\nu(z)$ denotes the parabolic cylinder function \cite{dlmf} and $a,b$ are two arbitrary constants independent of $ \phi$. Using the boundary conditions for $\phi \to\pm\infty$, and the fact that $\tilde{P}(k,\phi,s)$ is continuous at $\phi=0$  we have,
\bea
\tilde{P}(k,\phi,s)= \left \{
\begin{array}{ll}
     a~ D_{-q}\left(\phi \sqrt{\frac{2k}\Lambda}\right) , & \text{for  } \phi >0 \\
      a~ D_{-q}\left(-\phi \sqrt{\frac{2k}\Lambda}\right), & \text{for } \phi <0.
    \end{array} \right.
    \label{homosol}
\eea
Integrating Eq.~\eqref{eq:Ptdiff} across $\phi=0,$ we get, 
\bea
\frac{\id \tilde{P}}{\id \phi}\bigg|_{\phi=0^+}-\frac{\id \tilde{P}}{\id \phi}\bigg|_{\phi=0^-}=-\frac{\sin(k y_0)}{\Lambda^2}.\n
\eea
Using this equation with Eq.~\eqref{homosol} we get, 
\bea
a= \frac{2^{\frac q 2} \sin (k y_0)}{\sqrt{8 \pi k \Lambda^3}} \Gamma\left(\frac q 2 \right).
\label{eq:a}
\eea
Finally, combining Eq.~\eqref{eq:a} with Eq.~\eqref{homosol} we get,
\bea
\tilde{P}(k,\phi,s)&=& \frac{2^{\frac q 2} \sin(k y_0)}{\sqrt{8 \pi k \Lambda^3}} \Gamma\left(\frac q 2 \right) D_{-q}\left(|\phi| \sqrt{\frac{2k}\Lambda}\right),
\eea
where, as before, we have denoted $q= \frac 12(1+ \frac s{k \Lambda}).$ 
Since we are interested in the $y$-marginal distribution, we integrate over $\phi$ to get,
\bea
\hat{P}(k,s)=\frac{2 \sin(k y_0)}{s + k\Lambda} ~  _2F_1\left(1,\frac{q+1}2,\frac {q+2}2,-1\right), \n
\eea
which is the  $\sin$--Laplace transform of $P(y,t).$
Here $_2F_1(a,b,c,z)$ denotes the Hypergeometric function \cite{dlmf}. 

To find the position distribution we need to invert the  Laplace and $\sin$ transformations. The inverse Laplace transform is defined by the integral,
\bea
\hat P(k,t) = \int_{c_0-i \infty}^{c_0+ i \infty} \id s~ e^{st} \tilde{P}(k,s),
\eea
where $c_0$ is chosen such that all the singularities of the integrand lie to the left of the $Re(s)=c_0$ line. To compute the above integral let us first recast $\tilde P(k,s)$ as,
\bea
\tilde{P}(k,s) &=& \frac{2  \sin(k y_0)}{s + k\Lambda} ~  _2\tilde F_1\left(1,\frac{q+1}2,\frac {q+2}2,-1\right) \Gamma\left(\frac {q+2}2 \right), \label{eq:Pks}
\eea
where $_2\tilde F_1(a,b,c,z) = _2F_1(a,b,c,z)/\Gamma(c)$ denotes the regularized Hypergeometric function which is analytic for all values of $a,b,c$ and $z.$ From Eq.~\eqref{eq:Pks}, it is straightforward to identify the singularities of $\tilde{P}(k,s),$ on the complex $s$-plane all of which lie on the negative real $s$-axis: $s_n=-k \Lambda(4n+5)$ with $n=-1,0,1,2,\cdots$  where  $s_{-1}$ comes from the prefactor $(s+\Lambda k)^{-1}$ while $s_{n \ge 0}$ are obtained from the singularities $q_n=-2(n+1)$  of  $\Gamma\left(\frac{q+2}2 \right).$

The inverse Laplace transform of Eq.~\eqref{eq:Pks} can then be expressed as
\bea
\hat P(k,t) &=& \sum_{n=-1}^{\infty} e^{s_n  t} R_n, \label{eq:res_sum}
\eea
where $R_n$ denotes the  residue of $\tilde{P}(k,s)$ at $s=s_n.$ These residues can be computed exactly and turn out to be 
\bea
R_n =  2 \sin(k y_0) \frac{(-1)^{n+1}}{(n+1)!} ~_2\tilde F_1 \left(1,-n - \frac 12,-n,-1 \right). \n
\eea
Using the above expression in Eq.~\eqref{eq:res_sum} and shifting $n \to n-1$, we get,  
\bea
\hat P(k,t) &=& 2 \sin(k y_0) \sum_{n=0}^{\infty} \frac{(-1)^n}{n!}  e^{- (1+4 n) k \Lambda t}  2\tilde F_1 \left(1, -n+ \frac 12, -n+1,-1 \right). \label{eq:Pkt1}
\eea
Using properties of Hypergeometric functions, it can be shown that 
\bea
_2\tilde F_1\left(1, -n+ \frac 12, -n+1,-1 \right) = \frac {(-1)^n}{\sqrt 2} \left({-1/2 \atop n} \right) n!. ~~\n
\eea
Substituting the above identity in Eq.~\eqref{eq:Pkt1} we finally get,
\bea
\hat{P}(k,t)&=& \sqrt{2} \sin(k y_0) e^{-k \Lambda  t}\sum_{n=0}^{\infty} \left({- \frac 12 \atop n} \right) e^{- 4 n k \Lambda  t } 
= \frac{\sin(k y_0)}{\sqrt{\cosh \left(2 k  \Lambda t \right)}}. \label{eq:Linv}
\eea
The position distribution is given by the inverse $\sin$-transform,
\bea
P(y,t;y_0) =   \frac 2 \pi \int_{0}^\infty dk ~\sin (k y) \hat P(k,t) &=& \frac 1 \pi \int_{0}^\infty dk ~ \frac {2 \sin (k y) \sin (k y_0)}{\sqrt{\cosh \left(2 k  \Lambda t \right)}} \n \\ [0.25em]
&=& \frac 1 \pi \int_{0}^\infty dk ~ \frac {[\cos (k(y-y_0)) - \cos (k(y+y_0)) ]}{\sqrt{\cosh \left(2 k  \Lambda t \right)}}.\n
\eea
Clearly, $P(y,t)$ has a scaling form,
\bea
P(y,t;y_0) = \frac{1}{4\Lambda t}\Bigg[ f\Big(\frac{y-y_0}{4\Lambda t} \Big) - f\Big(\frac{y+y_0}{4\Lambda t}\Big) \Bigg],
\eea
where, the scaling function $f(z)$ can be evaluated exactly,
\bea
f(z) &=& \frac 1\pi \int_0^\infty d\kappa~ \frac{\cos(\kappa z)}{\sqrt{\cosh (\kappa/2)}} 
= \frac 1{\sqrt {2 \pi^3}} \Gamma \Big(\frac 14 + iz \Big)\Gamma \Big(\frac 14 - iz \Big).
\label{scaling_f}
\eea
The survival probability, given by Eq.~\eqref{eq:SP}, also has a scaling form,
\bea
S_y(t;y_0) = g\left(\frac {y_0}{4 \Lambda t}\right), \label{eq:S_scaling}
\eea
where $g(z_0)$ is given by,
\bea
g(z_0) = \int_0^\infty dz~ [f(z-z_0) -f(z+z_0)] = 2 \int_0^{z_0} dz~ f(z).
\eea
In terms of the original notation $t \to v_0^2 t /(2 \gamma)$, and $\Lambda^2=2\gamma D_R/v_0^2,$
\bea
S_y(t;y_0)=g\left(\frac {y_0}{v_0t}\sqrt{\frac \gamma{8 \dr}}\right).
\label{eq:S_scaling2}
\eea
The large time behavior can be extracted easily by taking $z_0 \ll 1,$
\bea
g(z_0)= 2 z_0 f(0) + O(z_0^2).
\eea
Thus, we have,
\bea
S(y_0,t)&\approx & \frac {\Gamma(1/4)^2}{2 \pi^{3/2}} \sqrt{\frac \gamma \dr} \, \frac {y_0}{v_0 t}\quad\text{for}\quad\frac{y_0}{v_0 t}\ll\sqrt{\frac{\dr}{\gamma}}\ll 1. 
\label{persis}
\eea
Using this result we conclude in the main text that the survival probability of a DRABP in the time regime (II) $\gamma^{-1}\ll t\ll D_R^{-1}$ has a power-law decay with persistence exponent $\alpha_y=1.$
\begin{figure}
\includegraphics[scale=0.6]{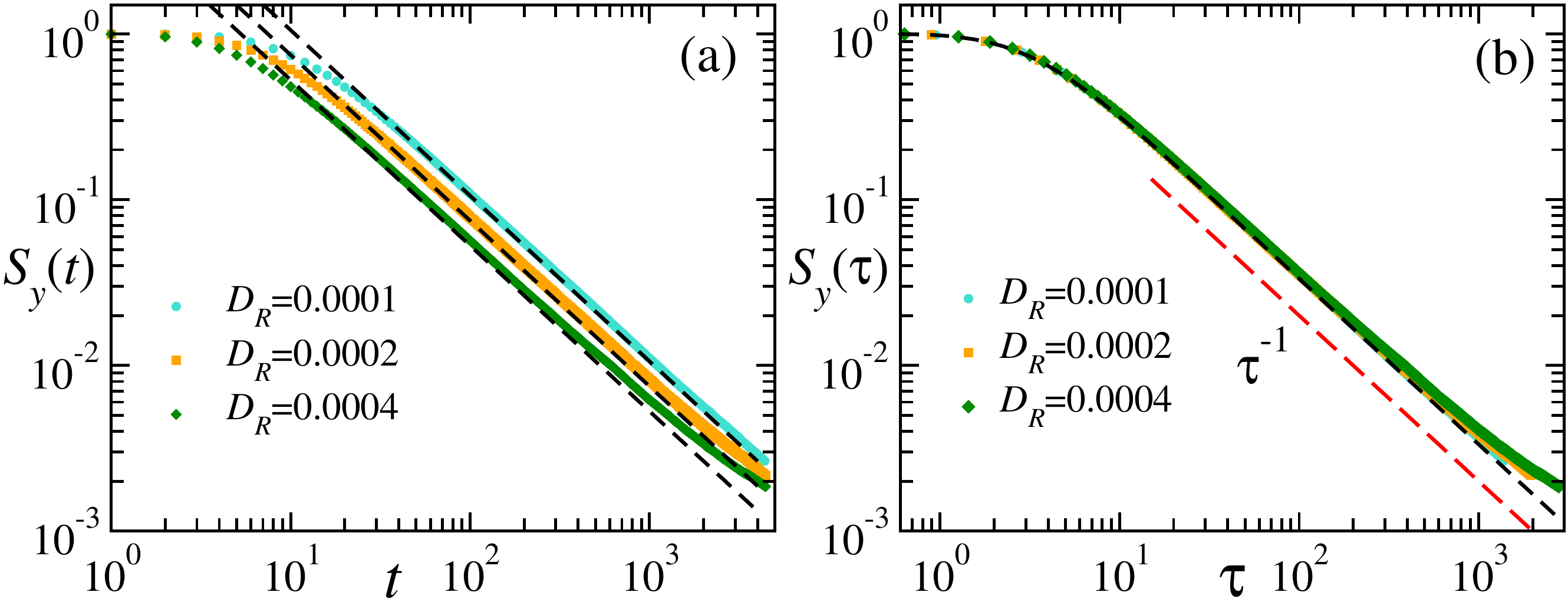}
\caption{(a) Survival probability $S_y(t;y_0)$ for $\gamma=80$ and three different values of $\dr$ starting from $y_0=0.01$. The symbols indicate the data obtained from numerical simulations and the dashed black lines indicate the analytical prediction in Eq.~\eqref{persis}.  (b) shows the same data plotted against the scaled time $\tau=v_0 t\sqrt{\frac{8\dr}{\gamma}}$, the dashed black line is obtained by numerically integrating Eq.~\eqref{surv_scaled}. The deviation from the analytical prediction at the tails is expected for $t\gtrsim \dr^{-1}$.}
\label{fig:surv_int}
\end{figure}

Note that, the exact first-passage distribution $F_y(t)= -\partial_t S_y(t;y_0)$ can be easily  computed from Eq.~\eqref{eq:S_scaling2},
\bea
F_y(t;y_0)=\frac{y_0\, \sqrt{2} \gamma^{3/2}}{v_0^3\, t^2\sqrt {\dr}} f\left(\frac {y_0}{v_0t}\sqrt{\frac \gamma{8 \dr}}\right).
\label{fpt_exact}
\eea
which was obtained in Ref.~\cite{Metzler} in the context of diffusing diffusivity. 
In terms of the scaled time $\tau=v_0 t\sqrt{\frac{8\dr}{\gamma}}$, 
\bea
S_y(\tau)=g(1/\tau)=2\int_{0}^{1/\tau}dz f(z)=2\int_{\tau}^{\infty}\frac{dz}{z^2}\, f(1/z),
\label{surv_scaled}
\eea
where $f(z)$ is defined in Eq.~\eqref{scaling_f}.

\section{Survival Probability along the direction of the initial orientation}
\begin{figure}[!h]
 \includegraphics[scale=0.4]{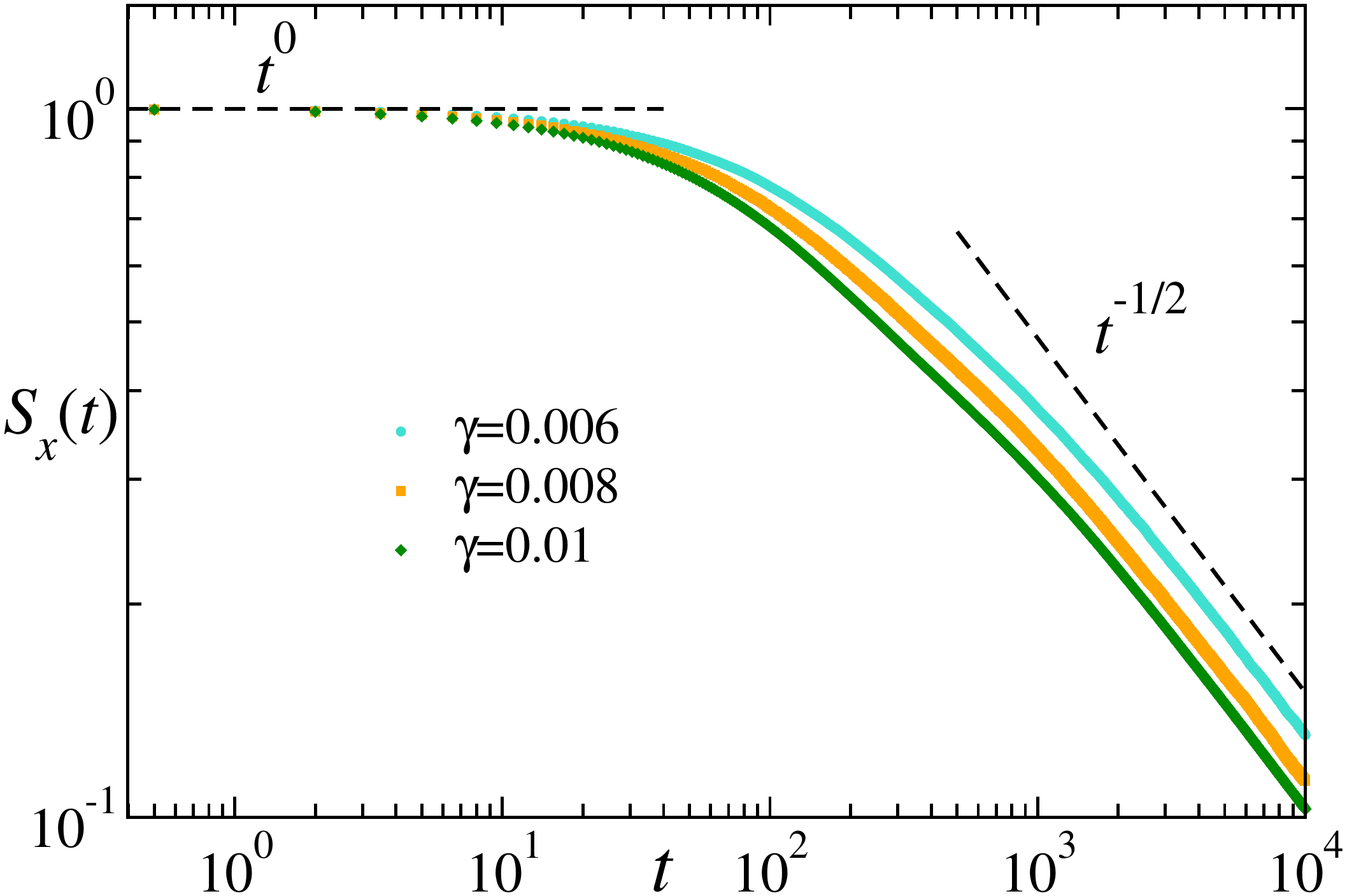}
 \caption{Survival probability $S_{x}(t;x_0)$ for three values of $\gamma$, $\dr=10^{-3}$ and $x_0=0.01$. The symbols denote the data obtained from numerical simulation. The black dashed lines indicate the analytical predictions.}
 \label{sx}
 \end{figure}

We are interested in the marginal survival probability along the direction of the initial orientation, i.e., the probability that a particle starting from $x_0>0$ with an initial orientation $\theta_0=0$ has not crossed the line $x=0$ up to time $t$.
 At very short times ($t\ll\gamma^{-1},\dr^{-1}$) the trajectories undergo none or very few reversals, as a result the particle starting with the initial orientation $\theta_0=0$ always moves away from the line $x=0$. Due to this the particle almost always survives and the corresponding survival probability remains unity. In the intermediate regimes (II) and (III) and large-time regime (IV) the particle exhibits diffusive motion as a result of which survival probability falls off as $t^{-1/2}$. We verify these predictions with numerical simulation in Fig.~\ref{sx}.
\newpage
\section{Movies}

We show the time-evolution of a single particle trajectory and of many non-interacting particles in two separate movies.

\begin{figure}[!h]
\centering
\textbf{\large Supplementary Movie 1: DRABP-trajectory.gif}
\vspace*{0.5 cm}

\fbox{\includegraphics[width=9cm]{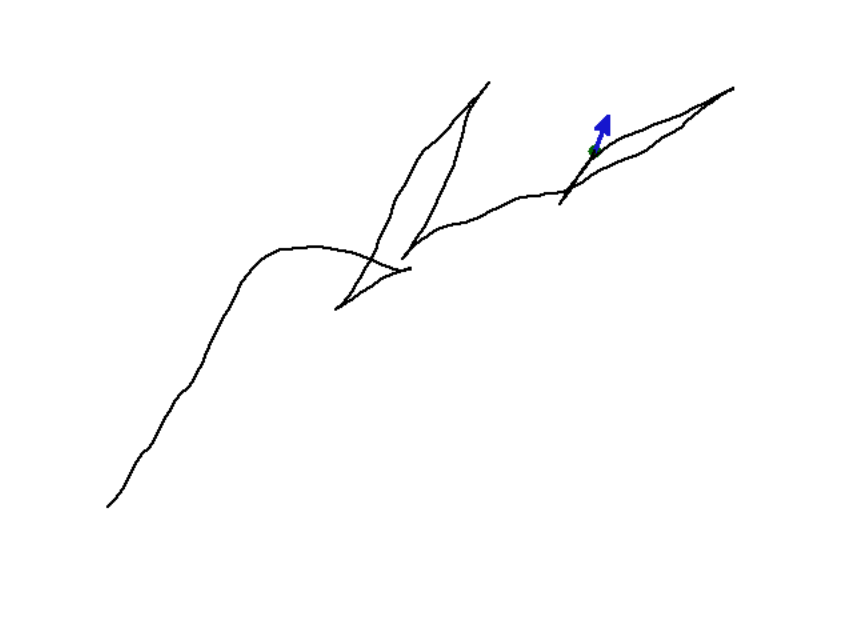}}
\caption*{Time-evolution of a DRABP starting from the origin with $\theta_0=\pi/4$ and $\sigma_0=1.$ Here $\dr=0.1$ and $\gamma=0.5$.}
\end{figure}

\begin{figure}[!h]
\centering
\textbf{\large Supplementary Movie 2: DRABP-profile1.gif}
\vspace*{0.5 cm}

\fbox{\includegraphics[width=9cm]{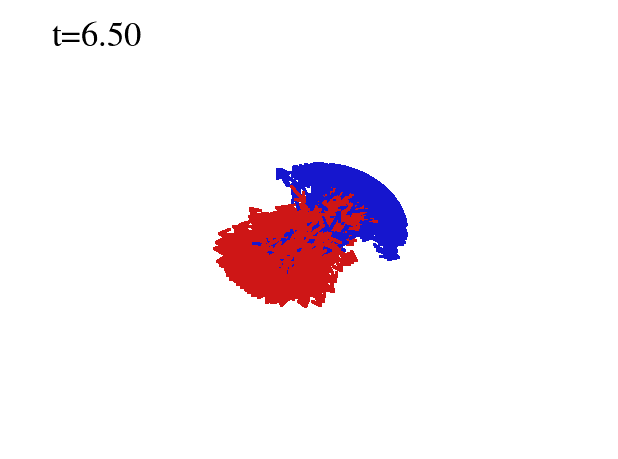}}
\caption*{Time-evolution of $N=5000$ non-interacting DRABP, each starting from the origin with $\theta_0=\pi/4$ and $\sigma_0=1.$ Here $\dr=0.02$ and $\gamma=0.1$.}
\end{figure}

\begin{figure}[!h]
\centering
\textbf{\large Supplementary Movie 3: DRABP-profile2.gif}

\vspace*{0.5 cm}
\fbox{\includegraphics[width=9cm]{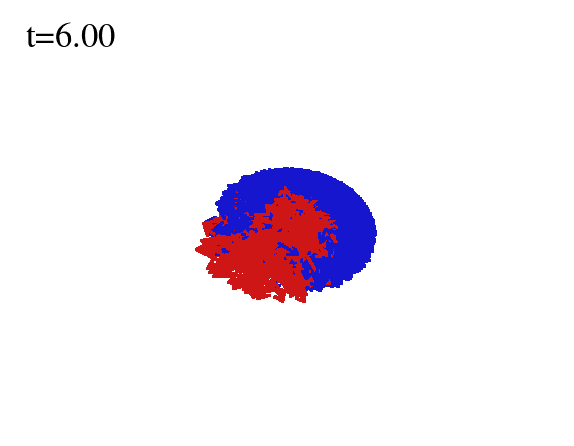}}
\caption*{Time-evolution of $N=5000$ non-interacting DRABP, each starting from the origin with $\theta_0=\pi/4$ and $\sigma_0=1.$ Here $\dr=0.1$ and $\gamma=0.02$.}\end{figure}